\documentclass[journal,draftcls,onecolumn,12pt,twoside]{IEEEtran}
\usepackage{graphicx} % Allows including images
\usepackage{booktabs} % Allows the use of \toprule, \midrule and \bottomrule in tables
\usepackage{amsmath}
\usepackage{amsfonts}
\usepackage{amssymb}
\usepackage{epstopdf}

\usepackage{dblfloatfix}
\usepackage[caption=false,font=footnotesize]{subfig}
\usepackage{color}
\usepackage{cite}

\begin{document}

\title{Multiuser Communication through Power Talk in DC MicroGrids}

\author{Marko~Angjelichinoski$^{\dagger}$,~\IEEEmembership{Student~Member,~IEEE,}
        \v Cedomir~Stefanovi\' c$^{\dagger}$,~\IEEEmembership{{Member,~IEEE},}
				Petar~Popovski$^{\dagger}$,~\IEEEmembership{Senior~Member,~IEEE,}
				Hongpeng~Liu$^{\ddagger}$,~\IEEEmembership{Member,~IEEE,}
				Poh~Chiang~Loh$^{\ast}$, Frede~Blaabjerg$^{\ast}$,~\IEEEmembership{{Fellow,~IEEE}}%
\thanks{$^{\dagger}$The authors are with the department of Electronic systems, Aalborg University, Denmark (e-mail: $\left\{\mbox{maa,cs,petarp}\right\}$@es.aau.dk).}
\thanks{$^{\ast}$The authors are with the department of Energy technology, Aalborg University, Denmark (e-mail: $\left\{\mbox{pcl,fbl}\right\}$@et.aau.dk).}
\thanks{$^{\ddagger}$The author is with the department of Electrical Engineering, Harbin Institute of Technology, Harbin, China‏ (e-mail: hpl@et.aau.dk).}
\thanks{The work presented in this paper was supported in part by EU, under grant agreement no. 607774 ``ADVANTAGE''.}}

\markboth{Journal of Selected Areas in Communications}%
{Angjelichinoski \MakeLowercase{\textit{et al.}}: Multiuser Power Talk for DC MicroGrids}

\maketitle

\begin{abstract}
Power talk is a novel concept for communication among control units in MicroGrids (MGs), carried out without a dedicated modem, but by using power electronics that interface the common bus.
The information is transmitted by modulating the parameters of the primary control, incurring subtle power deviations that can be detected by other units.
In this paper, we develop power talk communication strategies for DC MG systems with arbitrary number of control units that carry out all-to-all communication.
We investigate two multiple access strategies: 1) TDMA, where only one unit transmits at a time, and 2) full duplex, where all units transmit and receive simultaneously.
We introduce the notions of \emph{signaling space}, where the power talk symbol constellations are constructed, and \emph{detection space}, where the demodulation of the symbols is performed.
The proposed communication technique is challenged by the random changes of the bus parameters due to load variations in the system.
To this end, we employ a solution based on training sequences, which re-establishes the signaling and detection spaces and thus enables reliable information exchange.
The presented results show that power talk is an effective solution for reliable communication among units in DC MG systems.
\end{abstract}


\IEEEpeerreviewmaketitle

\section{Introduction}\label{Intro}

\IEEEPARstart{M}{icroGrids} (MGs) are localized clusters of Distributed Energy Resources (DERs) and loads interfacing one or multiple buses via flexible power electronic interfaces, able to operate both in connected (to the main grid) or standalone mode \cite{ref:1}.
The future power grid is envisioned as interconnected mesh of MGs, enabling flexible operation and improving the efficiency \cite{ref:2,ref:3}. Variety of practical MG applications have emerged, differing significantly in scope, size and in the demands for communication and signal processing support. The MG operation, especially in standalone mode \cite{ref:4}, relies on advanced control mechanisms assisted by communication technologies.
MG control is commonly based on a three-level hierarchy: primary, secondary and tertiary \cite{ref:5,ref:6,ref:7}.
The primary level provides fast control of the basic MG operation, such as bus voltage and/or frequency control based on predefined references.
The secondary and tertiary levels provide slower control mechanisms for enhancing the power quality, by setting the references for the primary level and optimizing the MG operation in grid-connected mode.

The traditional design of primary control avoids use of communications, due to the fact that most of the existing communication standards (particularly wireless ones), are not designed to support the machine-type control traffic in power grid applications, i.e., are unable to guarantee high reliability and constant availability.
Hence, the primary control is traditionally designed in a distributed manner, where each unit uses only locally available measurements \cite{ref:5,ref:6,ref:7}.
On the other hand, recent works consider the use of the existing MG power equipment, i.e., power electronics and power lines, as a means to exchange control messages \cite{ref:8,ref:9,ref:10,ref:11}.
An obvious solution in this respect is to  use power line communications (PLC) \cite{ref:12}.
However, for the emerging MG applications, where the focus is on small, isolated and localized systems operating in standalone mode, PLC might prove to be a cost-inefficient and overly complex solution.

This paper presents a novel, inexpensive, reliable and low-bandwidth communication solution, designed specifically for MGs operating in standalone mode and implemented using only the existing power equipment.
We refer to it as \emph{power talk}, as it modulates information over the main bus of the MG through subtle deviations of the power supplied by each unit.
In particular, power talk exploits the flexibility of the electronic inverters, which modulate the parameters of the primary loops that control the parameters of the common bus and in this way exchange information among the MG units.
The availability/reliability of power talk matches the ones of the MG bus, circumvents the use of additional hardware and requires only software enhancements.

Power talk has been introduced in \cite{ref:13,ref:14}, through a simplistic DC MG system with two units in a one-way communication scenario.
The focus of \cite{ref:13} is on enabling reliable communication without precise knowledge of the system configuration and of the load; it is shown that using a special input symbol in a role of a pilot transforms the MG bus into some of the well-studied channels.
\cite{ref:14} represents the unknown system configuration and load variations through a Thevenin equivalent, whose parameters determine the channel state that can be estimated.
This enables to design power talk constellations of an arbitrary order that perform optimally in terms of symbol error probability.

In this paper, we extend the power talk to scenarios with multiple units, in which each unit communicates with all other units in the system.
The contributions can be summarized as follows:
\begin{itemize}
\item We develop and analyze power talk strategies both for Time Division Multiple Access (TDMA), where a single unit transmits at a time, and Full Duplex (FD), where all units transmit and receive simultaneously.
%We show how to transform the MG bus into some well-studied channel models under certain mild assumptions for both communication strategies,
%We evaluate the strategies and outline their advantages and disadvantages.
\item We present the concepts of signaling and detection spaces, based on which we develop communication 
strategies. In particular, we design the signaling space by taking into account the control parameters that do not violate operational constraints of the MG.
We then investigate the detection space, i.e., the local set of voltages and currents that a unit can observe, based on which the demodulation is implemented.
\item We investigate simple protocol designs that deal with the problems of unknown system configuration and variable loads.
These protocols rely on using training sequences that reset the detection space at each receiver and foster reliable communications over the MG.
\end{itemize}

The rest of the paper is organized as follows.
Section \ref{sec:background} introduces the core ideas and implementation of power talk in DC MGs.
Section \ref{Basic} illustrates the communication principles of power talk through two simple, but insightful examples.
Section \ref{GeneralComm} presents the general communication model of power talk in DC MG.
Section \ref{Lim} investigates signaling and detection spaces and shows how to modulate and demodulate power talk symbols under MG operating constraints.
Section \ref{Protocols} addresses design aspects of fully operational communication protocols based on power talk.
Section \ref{perf} presents the performance evaluation.
Finally, Section \ref{Con} concludes the paper.

\section{Communication through Microgrid Control}
\label{sec:background}

\begin{figure*}[!t]
\centering
\subfloat[Primary control loops of VSC $k$ in standalone operation.]{\includegraphics[scale=0.27]{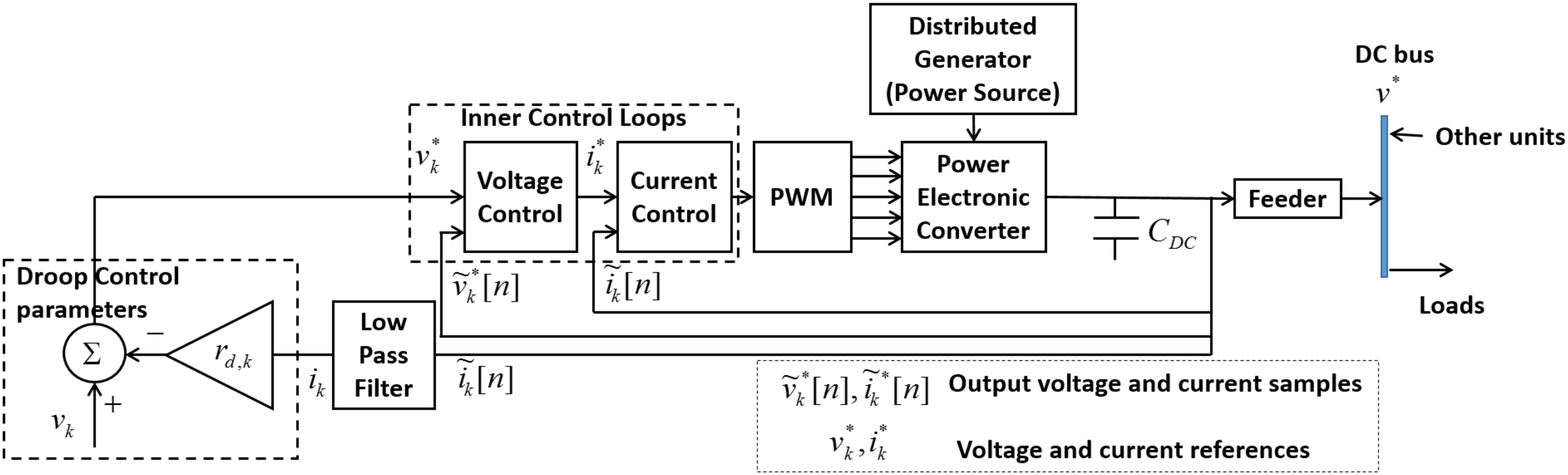}\label{control}}
\hfil
\subfloat[DC MG as communication system.]{\includegraphics[scale=0.27]{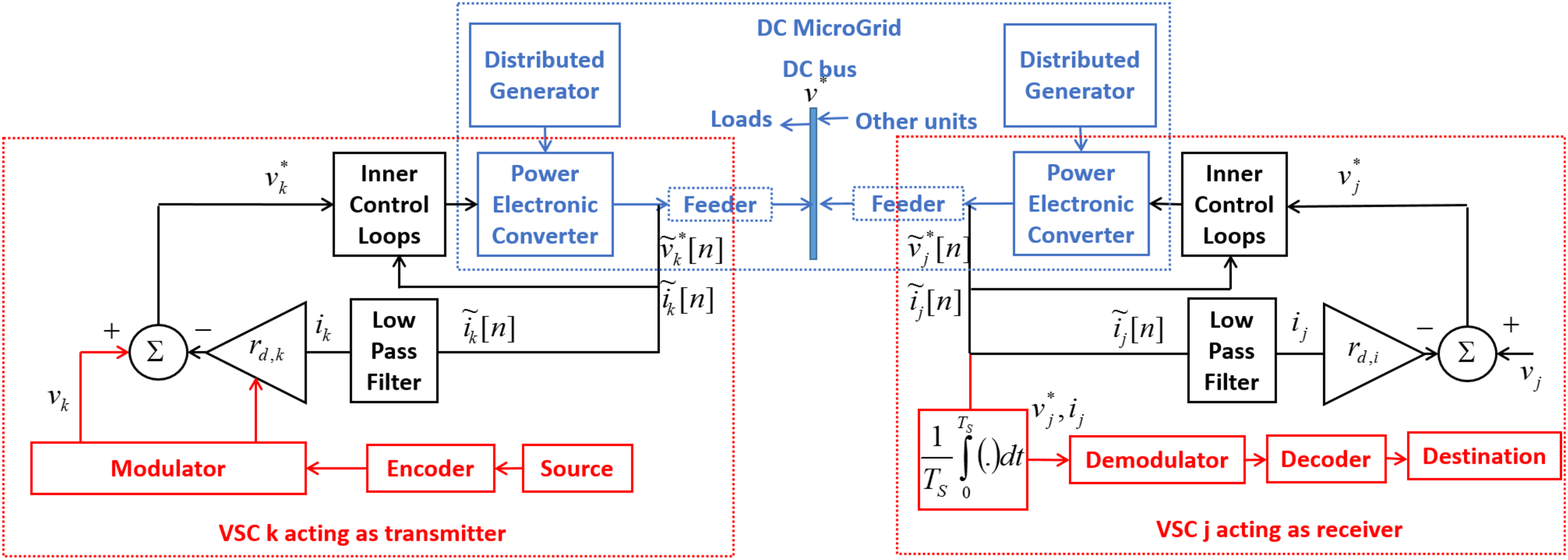}\label{MGCommDiagram}}
\caption{Overlaying communications over MG control.}
\label{MGContComm}
\end{figure*}

Fig.~\ref{MGContComm}\subref{control} shows the primary control diagram of a unit $k$ that operates as a Voltage Source Converter (VSC) and participates in the voltage regulation in the MG.\footnote{A unit can also operate as a Current Source Converter (CSC), when it does not use the inner voltage loop and does not participate in the voltage regulation.
Also, a unit can, in principle, switch between the VSC and CSC modes seamlessly; we consider only VSC units in the paper, as their task is to control the bus voltage and power sharing in the standalone mode.}
The unit interfaces the common bus through power electronics that implements the inner current and voltage control loops.
These loops are usually very fast (of the order of kHz), enforcing the output current $i_k$ and voltage $v_k^*$ to follow the predefined references, as shown on Fig.~\ref{MGContComm}\subref{control}.

The bus voltage reference $v_k^*$ that is fed to the inner voltage control loop of the unit $k$ is set according to the \emph{droop} law \cite{ref:15,ref:16,ref:17,ref:18}:
\begin{equation}\label{droop}
v_k^*=v_k - r_{d,k} i_k,
\end{equation}
where $v_k$ and $r_{d,k}$ are the droop parameters, namely the reference voltage and the virtual resistance (also referred to as droop slope), and $i_k$ is the output current.
Parameters $v_k$ and $r_{d,k}$ are controllable; in standalone mode their values are usually set to enable proportional power sharing based on the ratings of the units.
The basic idea of power talk is to change $v_k$ and $r_{d,k}$ in a controlled manner, thereby inducing variations in the output voltage $v_k^*$ and current $i_k$ that can be detected by other units in the system, thus leading to information exchange, see Fig.~\ref{MGContComm}\subref{MGCommDiagram}.

In the following, we list the assumptions used throughout the manuscript. Our focus is on small DC MGs, where all units are connected to a single common bus through feeder lines with negligible resistances. This assumption is valid for localized, isolated systems, expected to operate frequently in standalone mode \cite{ref:9,ref:10,ref:18}. For simplicity, we assume that VSCs in the MG supply a collection of loads aggregated in a single, purely resistive load, {denoted by $R$, whose instantaneous value is denoted by $r$.}
We note that same concepts can be used with minor modifications for mixture of loads, including constant power load. 
We consider all-to-all communication scenario in which the time is slotted, where VSC units in the system maintain slot synchronization.
The slot duration $T_s$ complies with the control bandwidth of the inner control loops and allows the system to reach steady-state \cite{ref:16}. In practice $T_s$ should be of the order of milliseconds. Each VSC samples the voltage and the output current with frequency $f_o$ of the order of kHz \cite{ref:17}. Power talk uses the averages of {$f_o T_s$ samples} during a single slot, i.e., all voltages and currents are averaged over slot duration $T_s$ with sampling frequency $f_o$, see Fig.~\ref{MGContComm}\subref{MGCommDiagram}. Finally, we consider only binary power talk and note that the developed techniques can be straightforwardly generalized to higher order modulations. 

Under the above assumptions, for general DC MG system with $K$ VSCs and each of them using \eqref{droop} to regulate output voltage and current, the common bus voltage $v^*$ in steady state is:
\begin{equation}\label{therip}
v^*=\frac{\sum_{k=1}^K\frac{v_k}{r_{d,k}+r_{l,k}}}{\frac{1}{r}+\sum_{k=1}^K\frac{1}{r_{d,k}+r_{l,k}}}\approx\frac{\sum_{k=1}^K\frac{v_k}{r_{d,k}}}{\frac{1}{r}+\sum_{k=1}^K\frac{1}{r_{d,k}}},
\end{equation}
where the approximation holds for negligible resistances of the feeder lines, $r_{l,1}\approx...\approx r_{l,K}\approx 0$.
The output current from VSC~$k$ in steady state is:
\begin{equation}\label{current}
i_k=\frac{v_k-v^*}{r_{d,k}},
\end{equation}
i.e., the output currents $i_k$ from each VSC are determined by the bus voltage.
Therefore, the bus voltage $v^*$ is the only degree of freedom that could be used for power talk in DC MGs.
Alternatively, this can be represented through a single parameter that is the output power $P_k=v^*i_k$ that VSC $k$ is supplying to the bus, leading to the term \emph{power talk}.
%Essentially, the units in power talk communicate though subtle deviations of the supplied power.

\begin{figure}[t]
\centering
\includegraphics[scale=0.22]{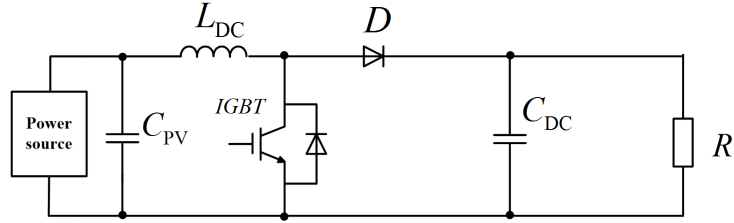}
\caption{DC-DC boost converter topology of a VSC unit.}
\label{boost}
\end{figure}

Finally, we note that all concepts, techniques and results presented in the paper were verified using PLECS\textregistered{} (Piecewise Linear Electrical Circuit Simulation) simulator integrated with Simulink\textregistered{} in realistic MG settings.
Specifically, we use VSC units in boost converter topology, see Fig.~\ref{boost}, interfacing DER to the main bus.
$IGBT$ represents the insulated-gate bipolar transistor with switching frequency $f_o$, $D$ is a fast recovery diode, $L_{DC}$ is the boost inductor and $C_{DC}$ is the output capacitor.
We simulate a low-voltage DC MG with allowable voltage deviation on the main bus $V_{min}\leq v^*\leq V_{max}$.
The maximum current rating of VSC $k$, related to the maximum output power of the respective DG is denoted by $I_{k,max}$.
In nominal mode, VSC $k$ operates with droop parameters $v_k^{\texttt{n}}$ and $r_{d,k}^{\texttt{n}}$, designed to enable proportional power sharing based on the rating of the unit, satisfying $r_{d,k}^{\texttt{n}}=\frac{v_k^{\texttt{n}}-V_{min}}{I_{k,max}}$.
The system is dimensioned to supply a collection of loads, equivalently represented with a single resistor $R$ whose instantaneous value varies in the range $r\in[R_{min},R_{max}]$.
The values of the parameters are summarized in Table~\ref{Param}.
The slot duration $T_s$ depends on the control bandwidth of the system.
The investigations performed using the simulated topology showed that for $T_s \geq 1 \, \text{ms}$, the system reaches a steady state.
Unless otherwise stated, in the rest of the text we use slot duration of $T_s =10 \, \text{ms}$.
%In summary, the above scenario represents small, localized DC MGs with DERs, in modular configuration.

\begin{table}
\caption{DC MG parameters simulated with PLECS.}
\label{Param}
\centering
\begin{tabular}{|c||c|c|c|c|c|c|c|c|c|c|}
\hline
Parameter & $ T_s $ & $f_o$ & $L_{DC}$ & $C_{PV}=C_{DC}$ & $V_{max}$ & $V_{min}$ & $I_{k,max}$ & $(v_k^{\texttt{n}},r_{d,k}^{\texttt{n}})$ & $R_{min}$ & $R_{max}$\\ \hline
Value & $ 10 \, \text{ms}  $ & $10 \, \text{kHz}$ & $5 \, \text{mH}$ & $470 \, \mu$F & $400 \, \text{V}$ & $390 \, \text{V} $ & $5 \, \text{A}$ & $(400 \, \text{V},2 \, \Omega)$ & $50 \, \Omega$ & $250 \, \Omega$\\ \hline 
\end{tabular}
\end{table}

\section{Two Illustrative Examples}
\label{Basic}

In this section, we present two toy examples of DC MGs, which capture the essence of power-talk concepts in all-to-all communication setup.

\subsection{DC MG with two units: TDMA and FD communication strategies}\label{Basic2}

\begin{figure}[h]
\centering
\includegraphics[scale=0.33]{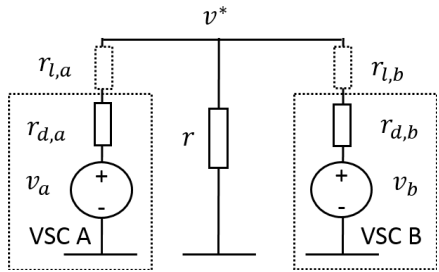}
\caption{DC MG system with two VSC units in steady-state.}
\label{2VSCMG}
\end{figure}

Fig.~\ref{2VSCMG} illustrates a two unit DC MG.
The VSCs maintain the common bus voltage $v^*$ through droop control and supply load $r$.
The droop control parameters of each VSC are denoted by $v_{a}$, $r_{d,a}$ and $v_b$, $r_{d,b}$.
As already noted, the feeder-line resistances are neglected, i.e., $r_{l,a}\approx r_{l,b}\approx 0$.
Each VSC locally observes the bus voltage $v^*$ and output current $i_k$,  $k \in \left\{a,b\right\}$.
Based on these observations, VSC $k$ constructs its local $v-i$ diagram as a the set of the outputs $(v^*,i_k)$, where each point in the diagram represents power $P_k=v^*i_k$ that VSC is supplying to the system.

The output power $P_k$ depends on: (i) the (own) droop parameters of VSC $k$, (ii) the droop parameters of the other VSC, and (iii) the value of the load $r$.
Assume that $r$ does not change during the slot.
Then, by measuring its own supplied power $P_k$, VSC $k$ can infer the droop parameters of the other unit.
Specifically, by tracking the point $(v^*,i_k)$ in the local $v-i$ diagram, each VSC determines how much power $P_k$ it is providing and implicitly learns the droop parameters of the other VSC.
This is the basic, underlying principle of power talk: each unit $k$ transmits information by changing its local droop control variables $v_k$ and $r_{d,k}$ and receives information by observing the local power output $P_k$ and detecting the corresponding point $(v^*,i_k)$ in the local $v-i$ diagram.
Henceforth, we refer to the $v-i$ diagram as the \emph{detection space}.

Denote the droop parameters of VSC $k$ as the input $\mathbf{x}_k$:
\begin{equation}\label{in_droop}
\mathbf{x}_k=(v_k,r_{d,k}),\; k\in\left\{a,b\right\}.
\end{equation}
In nominal mode of operation, when not ``power talking'', the droop parameters have values $\mathbf{x}_k^{\texttt{n}}=(v_k^{\texttt{n}},r_{d,k}^{\texttt{n}})$, $k\in\left\{a,b\right\}$, see Section \ref{sec:background}.
When transmitting, VSC $k$ uses two different combinations to represent the value of the transmitted bit $b_k$:
\begin{align}\label{in_binary_gen0}
\mathbf{x}_{k}^0=(v_{k}^0,r_{d,k}^0)\;\leftrightarrow\;``0",\\\label{in_binary_gen1}
\mathbf{x}_{k}^1=(v_{k}^1,r_{d,k}^1)\;\leftrightarrow\; ``1",
\end{align}
and we refer to combination $\mathbf{x}_{k}^{b_k}$ as an input symbol.
In the rest of the paper, we focus on the case where all units use the same symbols for signaling:
\begin{align}\label{in_binary}
\mathbf{x}_{k}^0\equiv\mathbf{x}^0,\;\mathbf{x}_{k}^1\equiv\mathbf{x}^1,\; k\in\left\{a,b\right\}.
\end{align}
This is the simplest case to deal with and provides valuable guidelines for designing power talk protocols.
%\textcolor{red}{Later}, we discuss generalizations of \eqref{in_binary}.
In principle, $\mathbf{x}^1$ and $\mathbf{x}^0$ can be chosen arbitrarily, as long as they comply to the operational constraints, as elaborated in Section~\ref{Lim}.

We denote the locally observed voltage and current at VSC $k$, i.e., the output symbol, by:
\begin{equation}\label{out_gen}
\mathbf{s}_k=(v^*,i_k),\; k\in\left\{a,b\right\}.
\end{equation}
Each value of $\mathbf{s}_k$ in the detection space is associated to output power $P_k = v^* i_k $ and depends on the values of droop parameters of both units.
In further text, we characterize $\mathbf{s}_k$ in the case of Time Division Multiple Access (TDMA) and Full Duplex (FD) approaches to power talk.

\subsubsection*{TDMA}
\begin{figure*}[!t]
\centering
\subfloat[VSC $a$ (the transmitter).]{\includegraphics[scale=0.55]{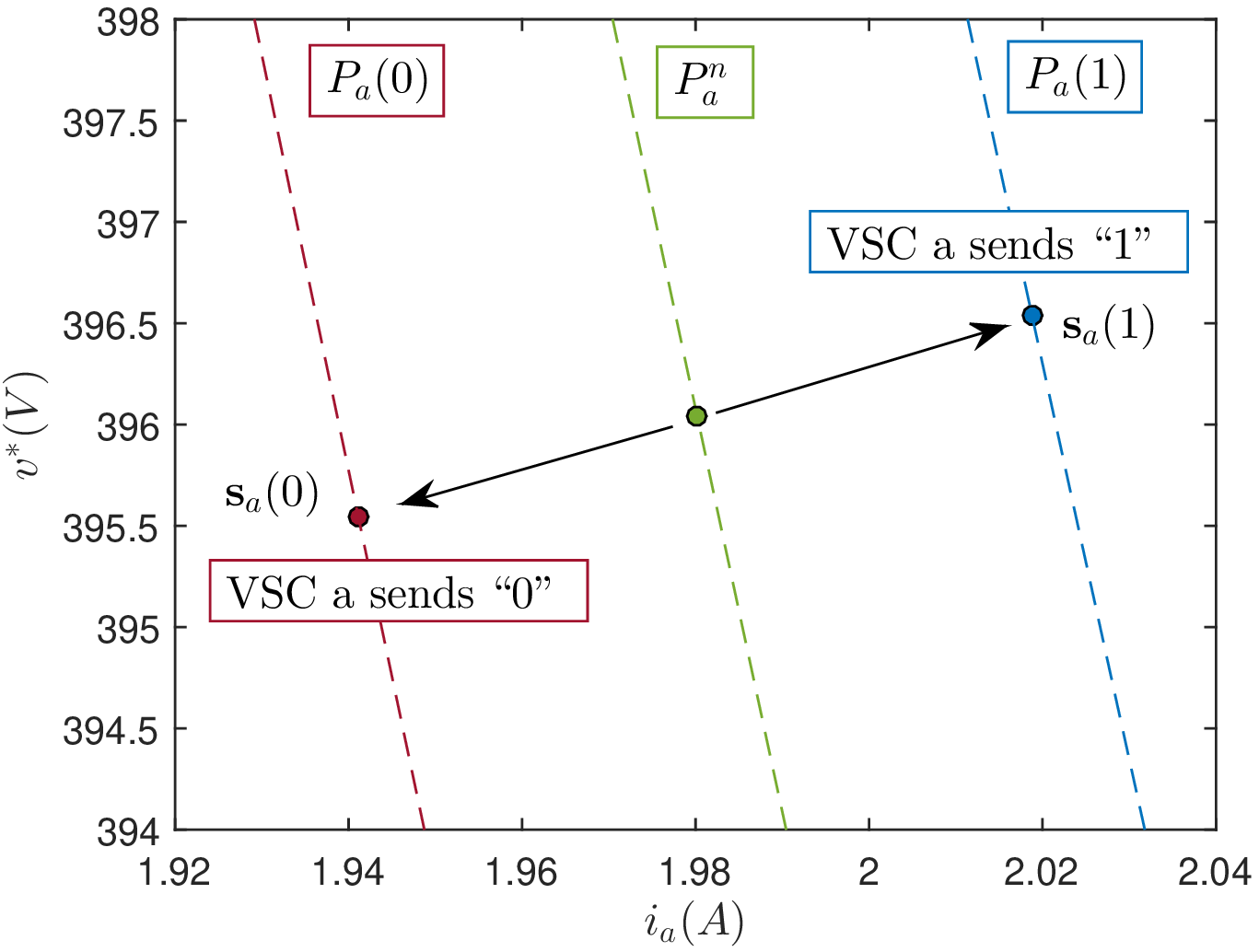}\label{2VSCTDMAA}}
\hfil
\subfloat[VSC $b$ (the receiver)]{\includegraphics[scale=0.55]{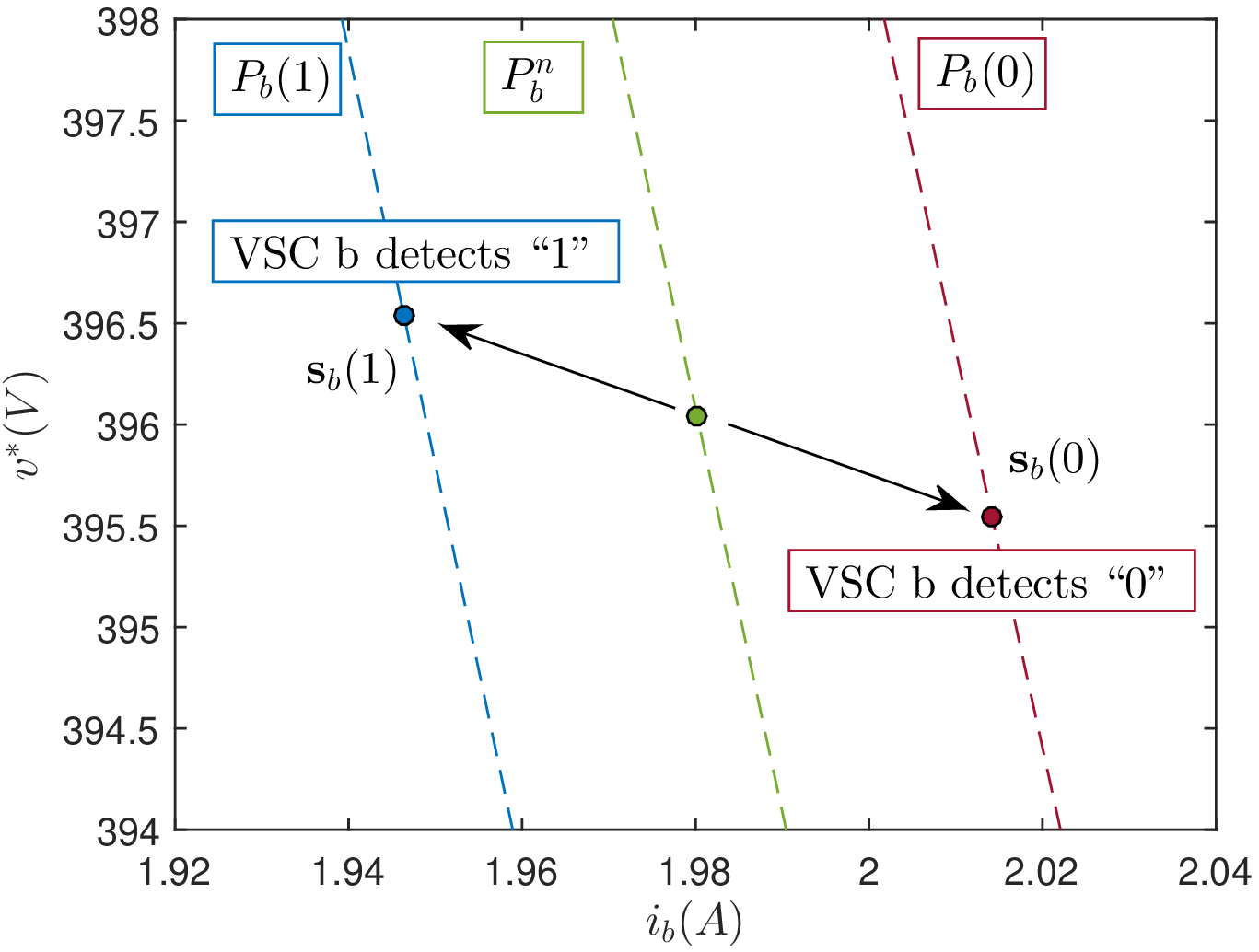}\label{2VSCTDMAB}}
\caption{$v-i$ diagram for TDMA-based binary power talk: 2 VSC units (obtained with PLECS\textregistered{} simulation of the system shown in Fig.~\ref{2VSCMG}, $v^0=399V,v^1=401V,r_d^0=r_d^1=2\Omega,v_a^{\texttt{n}}=v_b^{\texttt{n}}=400V,r_{d,a}^{\texttt{n}}=r_{d,b}^{\texttt{n}}=2\Omega$.)}
\label{2VSCTDMA}
\end{figure*}

Here we assume that the scheduling of the units is done in some predetermined manner, such that in each time slot only one active unit sends information over the MG, while the rest of the units operate in the nominal mode.
Assuming that $r$ does not change, it can be seen that the value of the output symbol at all units depends only on the value of the input symbol sent by the active unit VSC~$k$:
\begin{align}
\mathbf{s}_j = \mathbf{s}_j ( \mathbf{x}^{b_k} ) = \mathbf{s}_j ( b_k ) \text{ and } P_j = P_j ( \mathbf{x}^{b_k} ) = P_j ( b_k ), \; j \in \{ a, b \},
\end{align}
where $b_k$ is the information bit transmitted by VSC~$k$.
Without loss of generality, we assume that, when VSC $k$ is active, the inputs $\mathbf{x}^1$ and $\mathbf{x}^0$ satisfy:
\begin{align}
P_k ( 1 )> P_k^{\texttt{n}} > P_k( 0 ), \, k \in \left\{ a,b \right\},
\end{align}
where $P_k^{\texttt{n}}$ is the output power of VSC~$k$ when all VSCs operate in the nominal mode.
This implies that, if the load $r$ is stable, the output power of the receiving VSC $j$ satisfies:
\begin{align}
P_j ( 1 ) < P_j^{\texttt{n}} < P_j( 0 ), \, j \neq k.
\end{align}
Fig.~\ref{2VSCTDMA}\subref{2VSCTDMAA} illustrates detection spaces for VSCs in the example, assuming that VSC~$a$ is active. 
If VSC~$a$ inserts $\mathbf{x}^{0}$, it supplies less power than nominally, i.e., $P_a(0) < P_a^{\texttt{n}}$.
At the same time, VSC~$b$ observes $\mathbf{s}_b(0)$, VSC~$b$ detects that it supplies more than the nominal power $P_b(0)>P_b^{\texttt{n}}$, and concludes that VSC~$a$ is signaling ``0''.
Similarly, if VSC~$a$ inserts $\mathbf{x}^{1}$, VSC~$b$ observes $\mathbf{s}_b(1)$, detects that $P_b(1)<P_b^{\texttt{n}}$, and concludes that VSC~$a$ is signaling ``1''.
The TDMA scheme described above can be easily generalized to arbitrary number of units.

\subsubsection*{Full Duplex}
\begin{figure}[t]
\centering
\includegraphics[scale=0.55]{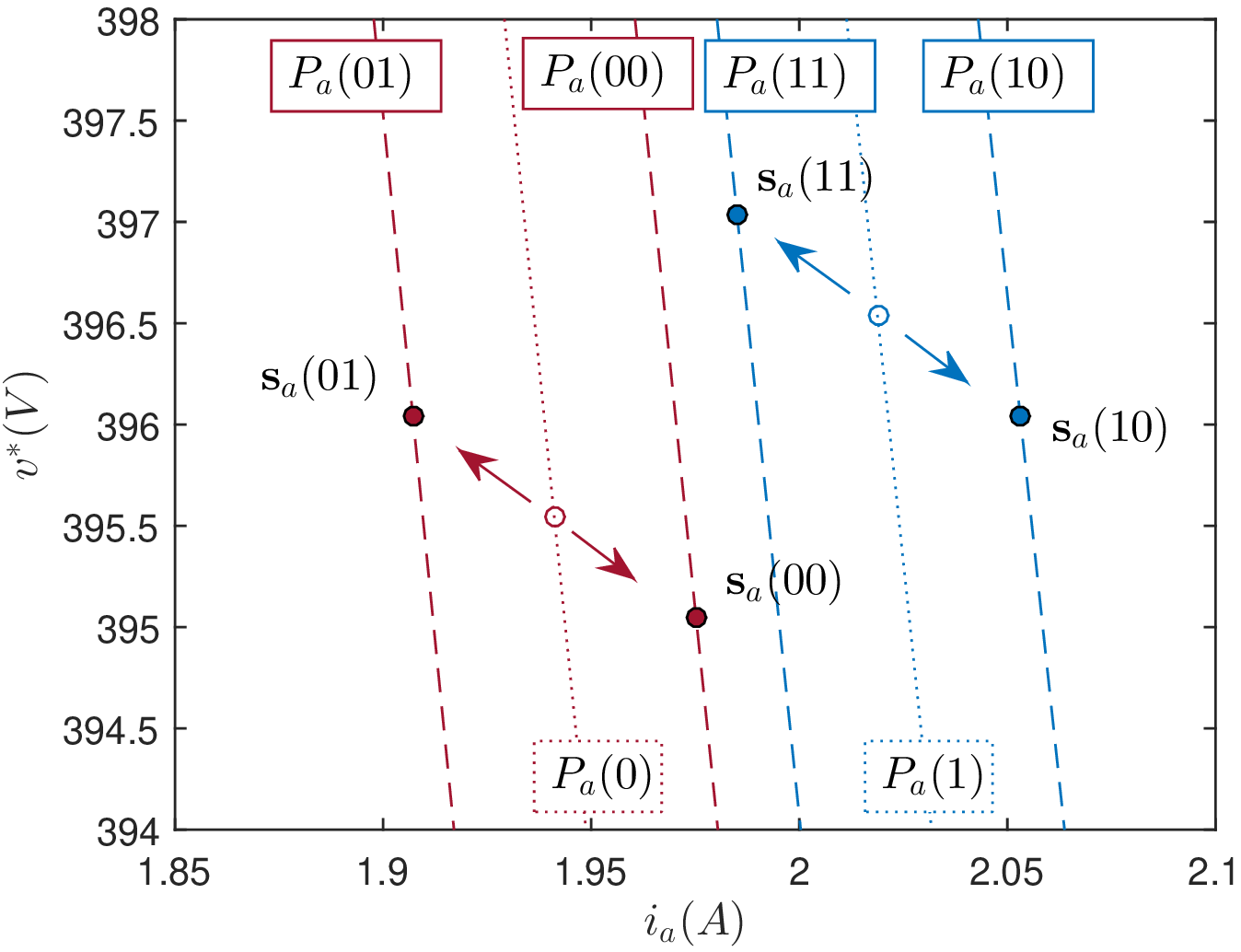}
\caption{$v-i$ diagram for FD-based binary power talk: 2 VSC units (obtained with PLECS\textregistered{} simulation of the system shown on Fig.~\ref{2VSCMG}, $v^0=399V,v^1=401V,r_d^0=r_d^1=2\Omega,v_a^{\texttt{n}}=v_b^{\texttt{n}}=400V,r_{d,a}^{\texttt{n}}=r_{d,b}^{\texttt{n}}=2\Omega$.)}
\label{2VSCMW}
\end{figure}

{In FD strategy, all units are simultaneously active.}
The same demodulation principle is applied: by observing its local output, a VSC can detect the information bit sent from the other VSC. 
The difference to the TDMA case is that the local output depends on the signaling combination of both VSCs:
\begin{align}
\mathbf{s}_k = \mathbf{s}_k ( b_a b_b ) \text{ and } P_k = P_k ( b_a b_b ), \; k \in \{ a, b \},
\end{align}
as illustrated on Fig.~\ref{2VSCMW} through the detection space for VSC $a$.
Consider the case when VSC~$a$ inserts $\mathbf{x}^{0}$, i.e., signals  $b_a=0$.
Then, depending on the symbol inserted by VSC~$b$, VSC~$a$ outputs different power $P_a$.
In particular, for $b_b = 0$, VSC~$b$ outputs less power than nominally, while for $b_b = 1$, VSC~$b$ outputs more power than nominally; correspondingly, $P_a$ increases or decreases.
The same reasoning applies when VSC~$a$ inserts $\mathbf{x}^{1}$, as well as for the detection of $b_a$ at VSC~$b$.

%\begin{figure*}[!t]
%\centering
%\subfloat[V-I diagram]{\includegraphics[scale=0.55]{2VSCMWDetS.eps}\label{2VSCMWDetS}}
%\hfil
%\subfloat[Histograms of observed voltage and current]{\includegraphics[scale=0.55]{2VSCMWVIout.eps}\label{2VSCMWVIout}}
%\caption{MW-based power talk for 2 VSC units: from VSC a perspective}
%\label{2VSCMW}
%\end{figure*}

The two VSC unit example exposes two crucial issues.
First, the described communication protocol requires prior knowledge of all possible points $\mathbf{s}_k$ in the detection space/local power outputs $P_k$.
As the detailed configuration of the MG is typically not known a priori, these values have to be learned in a predefined training phase, during which each VSC constructs the detection space.
Clearly, TDMA and FD strategies differ in terms of the amount of information required to construct the detection space. FD requires longer training phases and this is addressed in detail in Section~\ref{TrainCode}; here we only provide an illustration. In  system with two units, in TDMA binary power talk each unit has to learn two separate points when the other VSC unit transmits, such that the total number of points is four. 
In FD binary power talk, each unit has to learn four points in its detection space, leading to a total of eight points for the system. 
Second, the output power of a VSC can also vary as a result of the load change, which happens arbitrarily and randomly.
In particular, the current value of the load $r$ can be seen as a state of the system or the state of the communication channel.  
Whenever $r$ changes, the structure of the detection space also changes, leading to incorrect decisions at the receivers if the detection space prior to change is still used.
A strategy to deal with random state variations is to periodically repeat the training phase or to provide mechanism that tracks the state changes and re-initiates the training phase whenever a change is detected.
Section~\ref{Protocols} is dedicated to dealing with this challenge.

%We turn next to a three unit system, characterizing multiple access in MW power talk.

\subsection{DC MG system with three units: Characterization of Multiple Access}\label{Basic3}

\begin{figure}[h]
\centering
\includegraphics[scale=0.33]{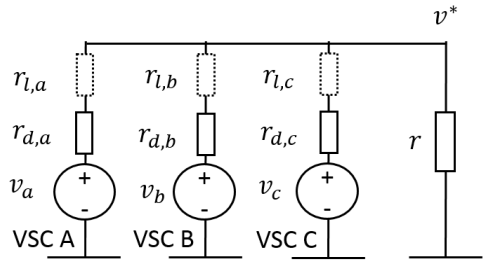}
\caption{DC MG system with three VSC units in steady-state.}
\label{3VSCMG}
\end{figure}

\begin{figure}[t]
\centering
\includegraphics[scale=0.55]{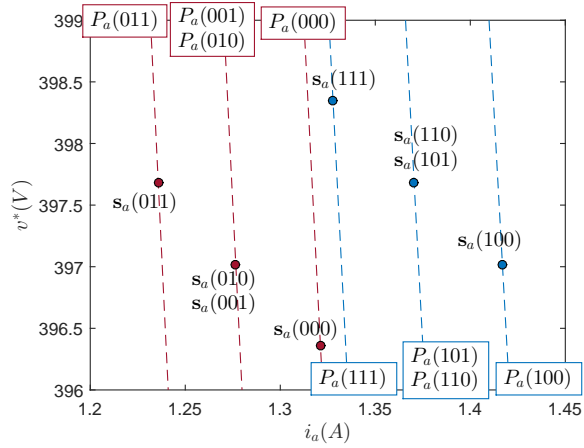}
\caption{$v-i$ diagram for FD-based binary power talk: 3 VSC units (obtained with PLECS\textregistered{} simulation of the system shown on Fig.~\ref{3VSCMG}, $v^0=399V,v^1=401V,r_d^0=r_d^1=2\Omega,v_a^{\texttt{n}}=v_b^{\texttt{n}}=v_c^{\texttt{n}}=400V,r_{d,a}^{\texttt{n}}=r_{d,b}^{\texttt{n}}=r_{d,c}^{\texttt{n}}=2\Omega$.)}
\label{3VSCMW}
\end{figure}

Consider a DC MG system with three VSC units, denoted with $a$, $b$ and $c$, see Fig.~\ref{3VSCMG}.
The units can communicate using TDMA or FD strategy.
The TDMA solution, outlined in the previous example, can be straightforwardly generalized to arbitrary number of units in the system and the basic communication principles remain the same as illustrated in Fig.~\ref{2VSCTDMA}.
The FD power talk, however, faces an additional challenge.
Assuming that the load is stable and applying the notation $\mathbf{s}_k = \mathbf{s}_k ( b_a b_b b_c )$ and $P_k = P_k ( b_a b_b b_c )$, VSC~$a$ can observe the following symbols in the detection space $s_a$/output powers $P_a$, as depicted in Fig.~\ref{3VSCMW}:
\begin{eqnarray}
\mathbf{s}_a(000) &\leftrightarrow& P_a(000); \qquad \mathbf{s}_a(011)\leftrightarrow P_a(011); \qquad \mathbf{s}_a(001)\cong\mathbf{s}_a(010)\leftrightarrow P_a(001)\cong P_a(010) \nonumber \\
\mathbf{s}_a(100) &\leftrightarrow& P_a(100); \qquad \mathbf{s}_a(111)\leftrightarrow P_a(111); \qquad \mathbf{s}_a(101)\cong\mathbf{s}_a(110)\leftrightarrow P_a(101)\cong P_a(110) \nonumber
\end{eqnarray}
Obviously, the outputs $\mathbf{s}_a(001)$ and $\mathbf{s}_a(010)$, as well as the outputs $\mathbf{s}_a(101)$ and $\mathbf{s}_a(110)$, are indistinguishable.
In other words, VSC $a$ can not distinguish between the cases in which the sum of the bits of the other units is the same, as then the sum of their output powers is the same and, thus, the output power $P_a$ is the same.
Also, it is easy to verify that the value of $\mathbf{s}_a/P_a$ depends on the bit signaled by VSC~$a$ and the integer sum, i.e., the \emph{Hamming weight}, of the bits signaled by the other units.
Summarizing, the FD binary power talk system, as seen from each VSC locally and given the value of the local input, can be equivalently represented by a Multiple Access Adder Channel with Binary Inputs (BI-MAAC).
%This, in turn, calls for the corresponding BI-MAAC coding methods in order to obtain individual bit streams from the aggregate observations, which is treated in Section~\ref{Protocols}.
{This example illustrates another major difference between TDMA and FD power talk.
Namely, in TDMA by demodulating the symbols in the detection space (see Fig.~\ref{2VSCTDMAB}) the information bit is directly obtained.
FD power talk (see Fig.~\ref{3VSCMW}), in turn, calls for the corresponding BI-MAAC coding methods in order to obtain individual bit streams from the aggregate observations, i.e., the modulation and coding with FD are separated.
Section \ref{Lim} introduces the detection mechanism used for symbol demodulation for TDMA and FD, whereas multiple access coding for FD is treated in Section~\ref{Protocols}.}

\section{General Binary Communication Model}\label{GeneralComm}

Consider a DC MG with $K$ units, connected in parallel to a common bus through feeder lines with negligible resistances, supplying resistive load $r$.
Assume a binary communication scheme, in which units send $\mathbf{x}^1=(v^1,r_{d}^1)$ for bit value ``1'' and $\mathbf{x}^0=(v^0,r_{d}^0)$ for bit value  ``0''.
Without loss of generality, assume that $P_k ( 1 ) > P_k( 0 ), \forall r\in[R_{min},R_{max}]$ where $P_k ( b_k )$ is the output power of VSC $k$ when transmitting bit $b_k$ and all other units operate in nominal mode.
The output of unit $k$ is  $\mathbf{s}_k=(v^*,i_k)$, the output consists of the bus voltage $v^*$ and output current $i_k$, which determine the output power $P_k$.
As noted above, $\mathbf{s}_k$ and, thus, $P_k$ depend on the inputs of all units, not just unit $k$, as well as on the value of the load $r$.
All units measure locally their outputs, i.e., unit $k$ observes its $\mathbf{s}_k$, based on which detection of the symbols of the active, signaling units is performed. 
The observation of $\mathbf{s}_k$, denoted by $\mathbf{y}_k=(\tilde{v}^*,\tilde{i}_k)$, differs from $\mathbf{s}_k=(v^*,i_k)$ due to voltage and current uncertainties caused by \cite{ref:19,ref:20,ref:21}: i) parasitic effects in the integrated circuits, ii) inaccuracies of the pulse width modulation process, iii) electromagnetic interference due to intentional/unintentional emissions by circuits in the system and from the surroundings, iv) ambient temperature, which is known to alter the performance of the power electronic components and v) sensor and measurement noise.
As suggested in \cite{ref:22,ref:23} and references therein, the aggregate uncertainty of the measurements in the feedback loops of the primary control, after averaging with LPF, can be accurately modeled with a Gaussian noise:
\begin{equation}\label{out}
\mathbf{y}_k=(\tilde{v}^*,\tilde{i}_k)=\mathbf{s}_k+\mathbf{z}_k,
\end{equation}
where $\mathbf{z}_k\sim\mathcal{N}(\mathbf{0},\text{diag}\left\{\sigma_{v^*}^2,\sigma_{i_k}^2\right\})$.
Therefore, the observation $\mathbf{y}_k$, given {the momentary value of the load $r$}, will be distributed according to the Gaussian conditional pdf $p(\mathbf{y}_k|\mathbf{s}_k,r)$.

\begin{figure}[t]
\centering
\includegraphics[scale=0.3]{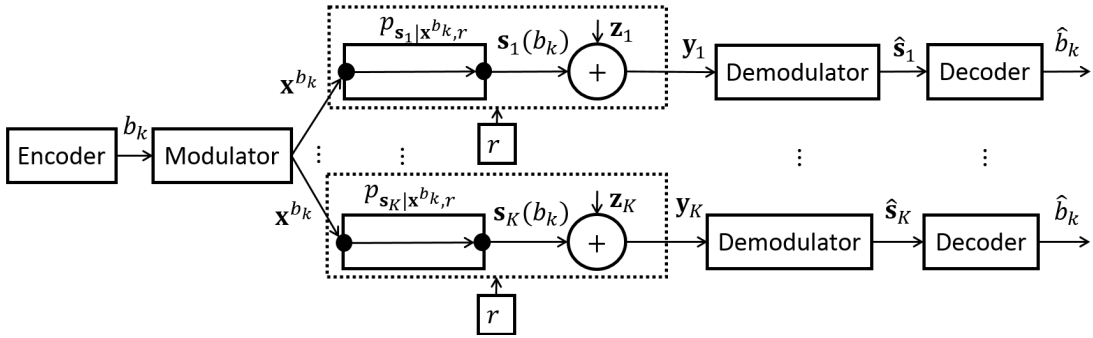}
\caption{Communication model for TDMA-based binary power talk: {VSC $k$ as a transmitter}.}
\label{TDMACommGen}
\end{figure}

For TDMA power talk, a unit transmits only in time slots exclusively dedicated to it, while the rest of the units operate in the nominal mode.
Therefore, in every time slot, the MG can be equivalently represented as a broadcast channel from the signaling unit to all other units, as depicted on Fig.~\ref{TDMACommGen}.
When VSC $k$ transmits bit $b_k$, its input symbol $\mathbf{x}^{b_k}$ maps to output $\mathbf{s}_j ( b_k )$ at VSC $j$, $j = 1,\dots,K$.
In turn, VSC $j$ observes $\mathbf{y}_j = \mathbf{s}_j( b_k ) + \mathbf{z}$, and makes decision $\hat{b}_k$.

\begin{figure}[t]
\centering
\includegraphics[scale=0.3]{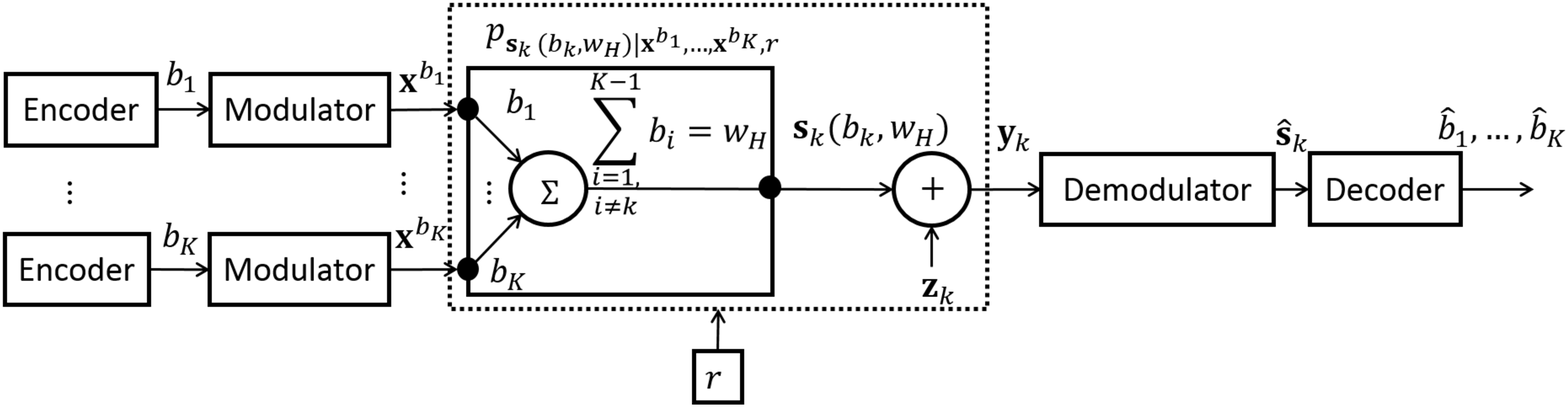}
\caption{General Communication model for FD-based binary power talk: {VSC $k$ as a receiver.}}
\label{MWCommGen}
\end{figure}

For FD power talk, all units transmit in every time slot, and the local observation of the output $\mathbf{y}_k$, $k = 1,\dots,K$, depends on the information bits of all units.
As illustrated in Section~\ref{Basic3}, the equivalent channel model, seen from each unit locally, can be represented by BI-MAAC with $K-1$ users, see Fig.~\ref{MWCommGen}.
The local output $\mathbf{s}_k$ depends on the combination of $b_k$ and $w_H ( \mathbf{b}_{\sim k} )$, where $w_H$ denotes Hamming weight and $\mathbf{b}_{\sim k}$ is the sequence of bits of all units except VSC $k$:
\begin{align}\label{FD_sk}
\mathbf{s}_k = \mathbf{s}_k ( b_k, w_H ( \mathbf{b}_{\sim k} ) ) \text{ and } P_k = P_k ( b_k, w_H ( \mathbf{b}_{\sim k} ) ), \, k =1, \dots, K,. 
\end{align}
The number of outputs that VSC $k$ may observe locally is $2K$, i.e., $K$ points for each value of the local bit $b_k$.
Also, for given $r$, it can be shown that the output powers at VSC $k$, $k = 1,\dots, K$ satisfy $P_{k}(b_k, W_1 ) > P_{k}(b_k, W_2 )$ if $W_1 < W_2 $, and where $W_1,W_2 = 0, \dots, K-1$ are Hamming weights, see \eqref{FD_sk}.

%Finally, both for TDMA and MW power talk, the effect of the observation noise can be captured with Discrete Memoryless Channel (DMC), whose cross over probabilities depend on the channel state $r$.
%In the following section, we focus on the design of symbol constellations and detection spaces as well as characterization of the corresponding DMC models.

\section{Communication under Constraints: Signaling and Detection Spaces}\label{Lim}

Here we introduce the concept of signaling space to capture the effect that the operating constraints of the MG have on the power talk schemes. 
%and to construct symbol constellations that do not violate these constraints.
We then investigate the structure of the detection space and its dynamics due to the load changes, as this is vital for implementing reliable power talk solutions.

\subsection{The signaling space}\label{Sspace}

Every MG is subject to operational constraints that may not be violated.
Denote by $\mathcal{C}$ a set that consists of all operational constraints.
{The signaling space $\mathcal{X}$} is the set of all possible symbols $\mathbf{x}_k$, $k=1,\dots,K$, that jointly satisfy the constraints in $\mathcal{C}$ for any value $r\in[R_{min},R_{max}]$:
\begin{equation}\label{Sspace_gen}
\mathcal{X}=\left\{\mathbf{x}_k=(v_k,r_{d,k}),\; k=1,...,K:\mathcal{C}, \;  r\in[R_{min},R_{max}]\right\},
\end{equation}
In general, $\mathcal{C}$ comprises constraints associated with system stability and power delivery quality, e.g., limits on the voltage, current, power dissipation, droop slope etc.
In this paper, we focus on the most important constraints, namely the bus voltage and output current constraints:
\begin{equation}\label{Cspace}
\mathcal{C}=\left\{V_{min} \leq v^* \leq V_{max}; I_{k,min} \leq i_k \leq I_{k,max}, k=1, \dots, K\right\},
\end{equation}
where $V_{min}$ and $V_{max}$ are the minimum and maximum allowable bus voltages and where $I_{k,min}$ and $I_{k,max}$ are the minimum and maximum output currents of VSC $k$; usually, $I_{k,min}=0$ and $I_{k,max}$ is the current rating of the unit.

{In this paper, we deal with binary power talk when all units employ the same symbols, i.e.,  $\mathbf{x}_k^1=\mathbf{x}^1=(v^1,r_{d}^1)$ and $\mathbf{x}_k^0=\mathbf{x}^0=(v^0,r_{d}^0)$.
%chosen such that $P_k(1)>P_k(0)$, $k=1,\dots,K$ without loss of generality.
Then, in the TDMA case, under constraints \eqref{Cspace} and the steady-state model \eqref{therip} and \eqref{current}, the symbols $\mathbf{x}^1$ and $\mathbf{x}^0$ should satisfy:}% \PP{Instead of $0/1$ you can use $b$ and then write after the equation that $b \in \{ 0,1\}$}
\begin{align}\label{Sspace_tdma_v}
r_{d}^{b}\frac{V_{min}-\frac{\sum_{i\neq k}\frac{v_i^{\texttt{n}}}{r_{d,i}^{\texttt{n}}}}{\big(\frac{1}{R_{min}}+\sum_{i\neq k}\frac{1}{r_{d,j}^{\texttt{n}}}\big)}}{\big(\frac{1}{R_{min}}+\sum_{i\neq k}\frac{1}{r_{d,i}^{\texttt{n}}}\big)^{-1}}+V_{min}\leq v^{b}\leq r_{d}^{b}\frac{V_{max}-\frac{\sum_{i\neq k}\frac{v_i^{\texttt{n}}}{r_{d,i}^{\texttt{n}}}}{\big(\frac{1}{R_{max}}+\sum_{i\neq k}\frac{1}{r_{d,i}^{\texttt{n}}}\big)}}{\big(\frac{1}{R_{max}}+\sum_{i\neq k}\frac{1}{r_{d,i}^{\texttt{n}}}\big)^{-1}}+V_{max},\\\label{Sspace_tdma_i}
\frac{\sum_{i\neq k}\frac{v_i^{\texttt{n}}}{r_{d,i}^{\texttt{n}}}}{\big(\frac{1}{R_{max}}+\sum_{i\neq k}\frac{1}{r_{d,i}^{\texttt{n}}}\big)}\leq v^{b}\leq r_{d}^{b}I_{k,max}+\frac{I_{k,max}}{\big(\frac{1}{R_{min}}+\sum_{i\neq k}\frac{1}{r_{d,i}^{\texttt{n}}}\big)}+\frac{\sum_{i\neq k}\frac{v_i^{\texttt{n}}}{r_{d,i}^{\texttt{n}}}}{\big(\frac{1}{R_{min}}+\sum_{i\neq k}\frac{1}{r_{d,i}^{\texttt{n}}}\big)},
\end{align}
for $k = 1, \dots, K$ and $b \in \{ 0,1\}$.
In FD case, all units simultaneously change their droop parameters and it can be shown that $\mathbf{x}^1$ and $\mathbf{x}^0$ should satisfy ($k = 1, \dots, K$ and $b \in \{ 0,1\}$):
\begin{align}\label{Sspace_mw_1}
r_{d}^b\frac{V_{min}}{KR_{min}}+V_{min} & \leq v^b\leq r_{d}^b\frac{V_{max}}{KR_{max}}+V_{max},\\
\frac{(K-1)\frac{v^0}{r_{d}^0}}{\big(\frac{1}{R_{min}}+\frac{K-1}{r_{d}^0}\big)} & \leq v^1 \leq I_{k,max}r_{d}^1+\frac{I_{k,max}}{\big(\frac{1}{R_{min}}+\frac{K-1}{r_{d}^0}\big)}+\frac{(K-1)\frac{v^0}{r_{d}^0}}{\big(\frac{1}{R_{min}}+\frac{K-1}{r_{d}^0}\big)},\\
\label{Sspace_mw_0}
\frac{(K-1)\frac{v^1}{r_{d}^1}}{\big(\frac{1}{R_{max}}+\frac{K-1}{r_{d}^1}\big)} & \leq v^0 \leq I_{k,max}r_{d}^0+\frac{I_{k,max}}{\big(\frac{1}{R_{max}}+\frac{K-1}{r_{d}^1}\big)}+\frac{(K-1)\frac{v^1}{r_{d}^1}}{\big(\frac{1}{R_{max}}+\frac{K-1}{r_{d}^1}\big)}.
\end{align}

\begin{figure*}[!t]
\centering
\subfloat[TDMA: $K=2$]{\includegraphics[scale=0.33]{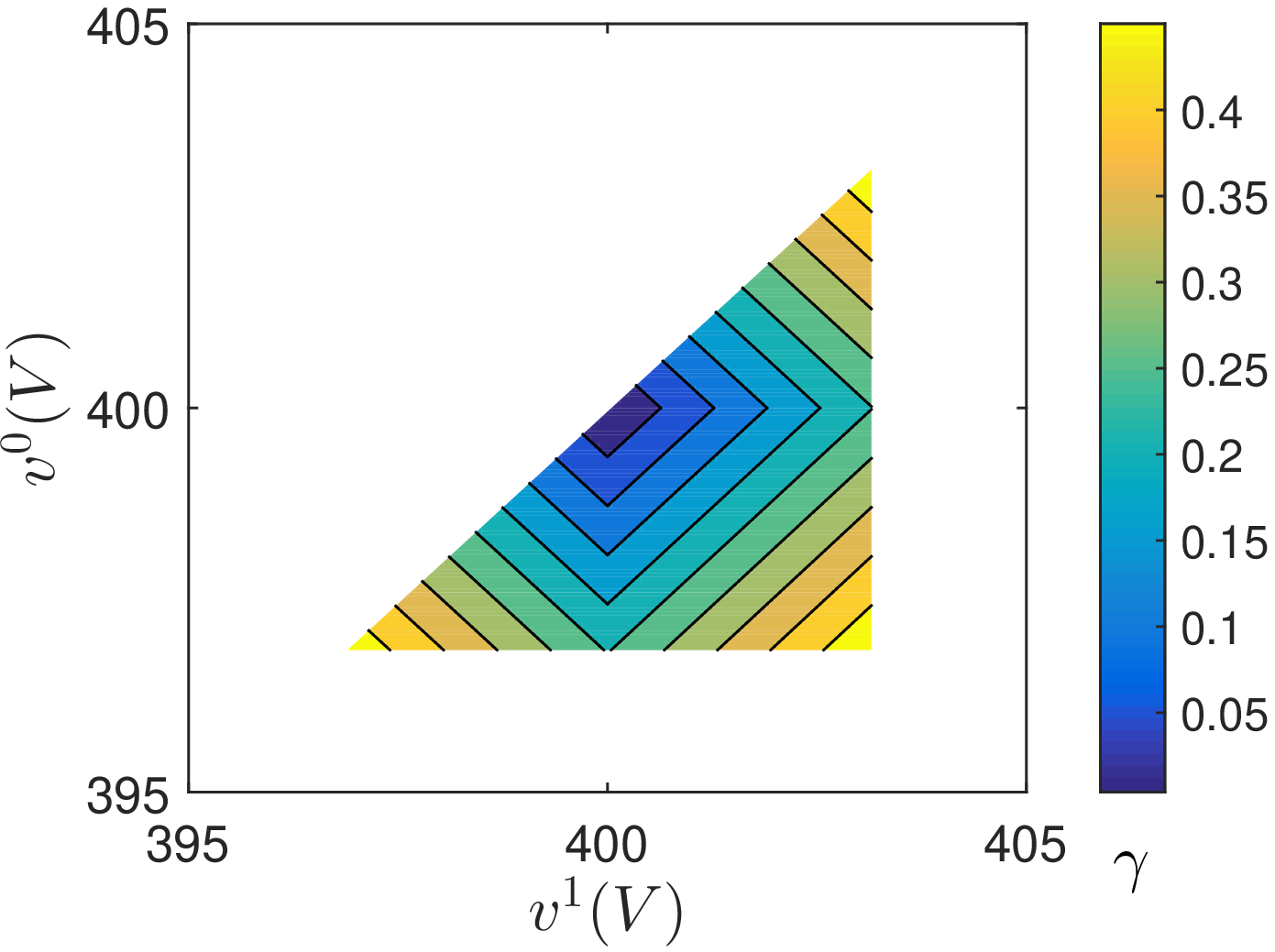}\label{SSpaceK2tdma}}
\hfil
\subfloat[TDMA: $K=3$]{\includegraphics[scale=0.33]{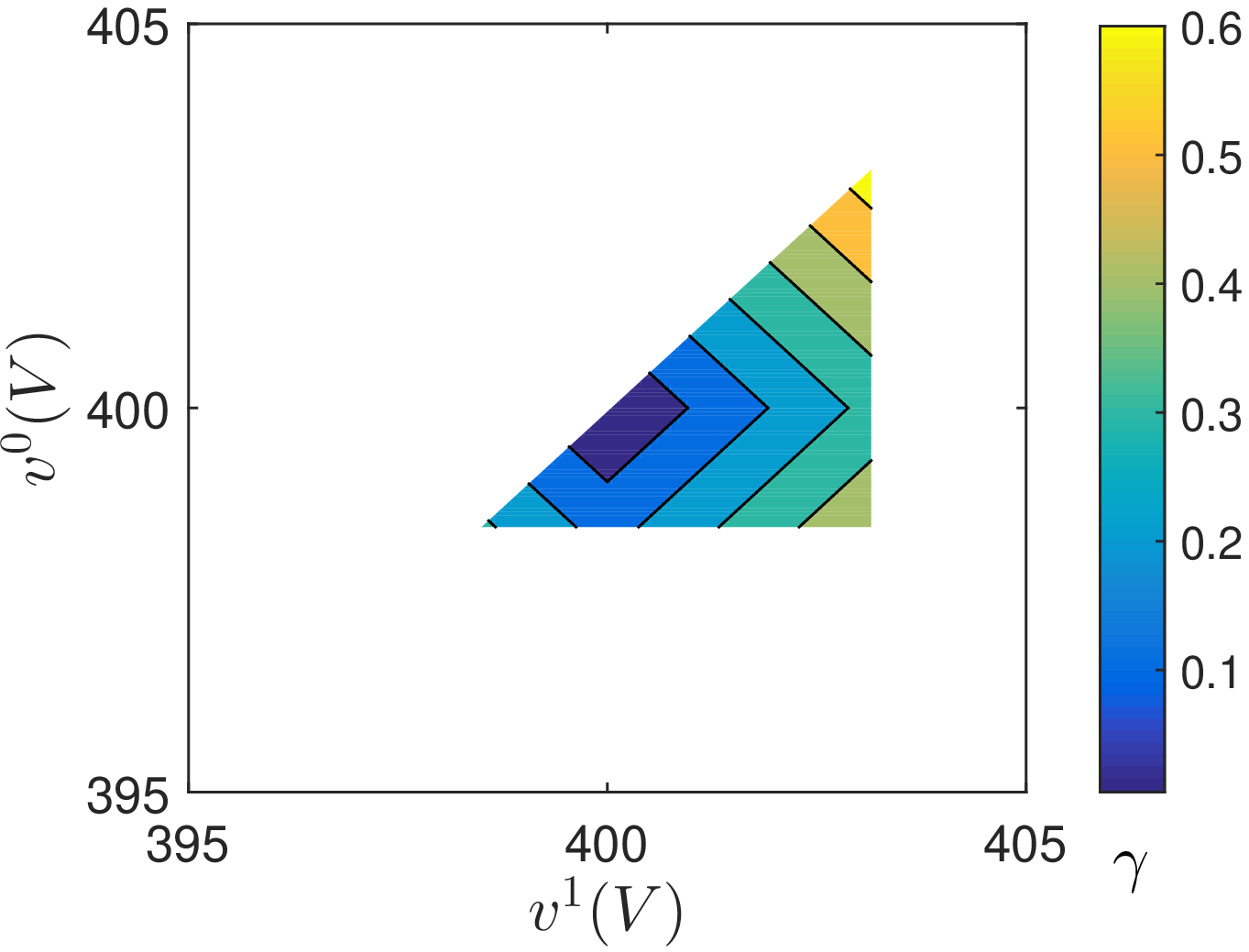}\label{SSpaceK3tdma}}
\hfil
\subfloat[TDMA: $K=4$]{\includegraphics[scale=0.33]{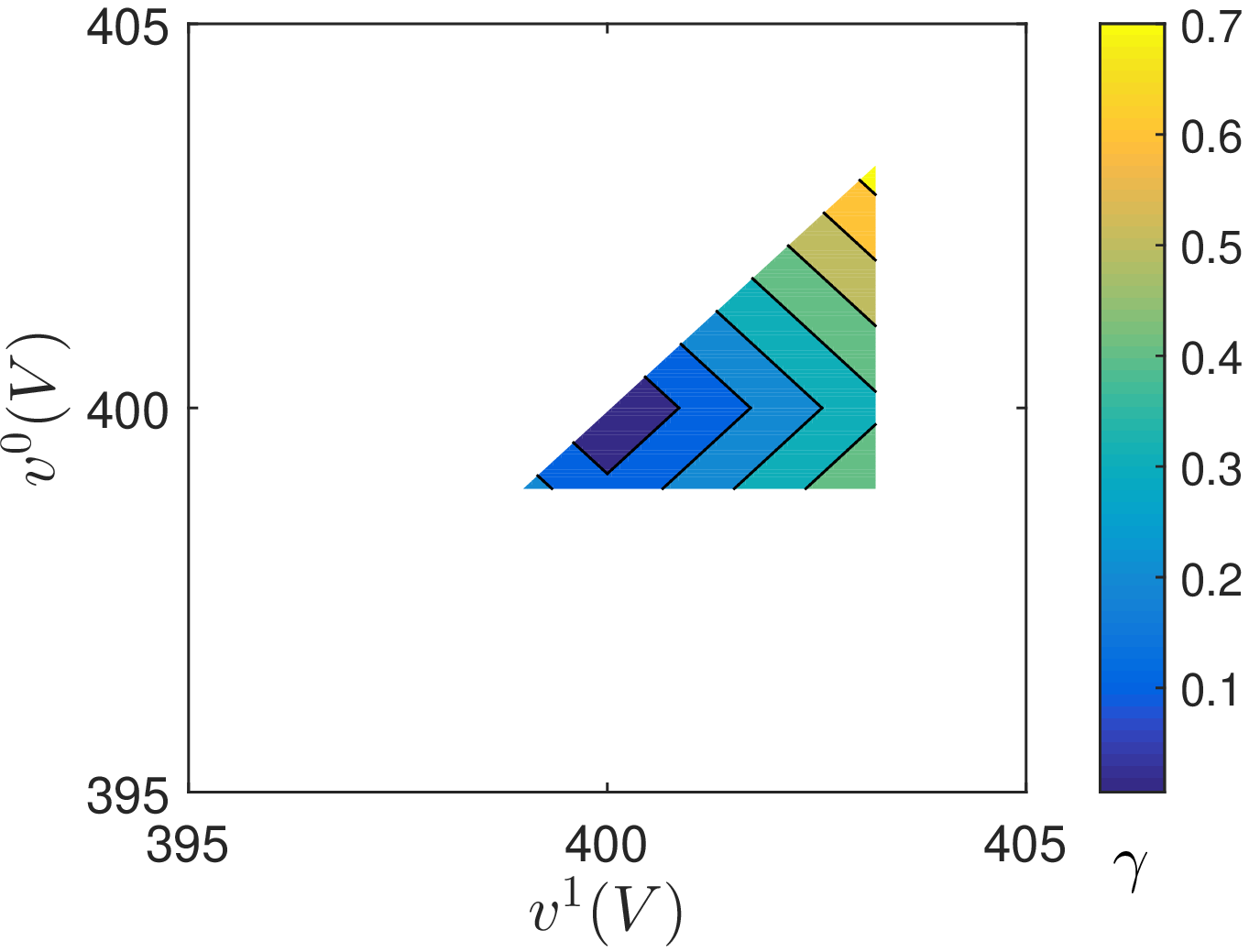}\label{SSpaceK4tdma}}
\hfil
\subfloat[FD: $K=2$]{\includegraphics[scale=0.33]{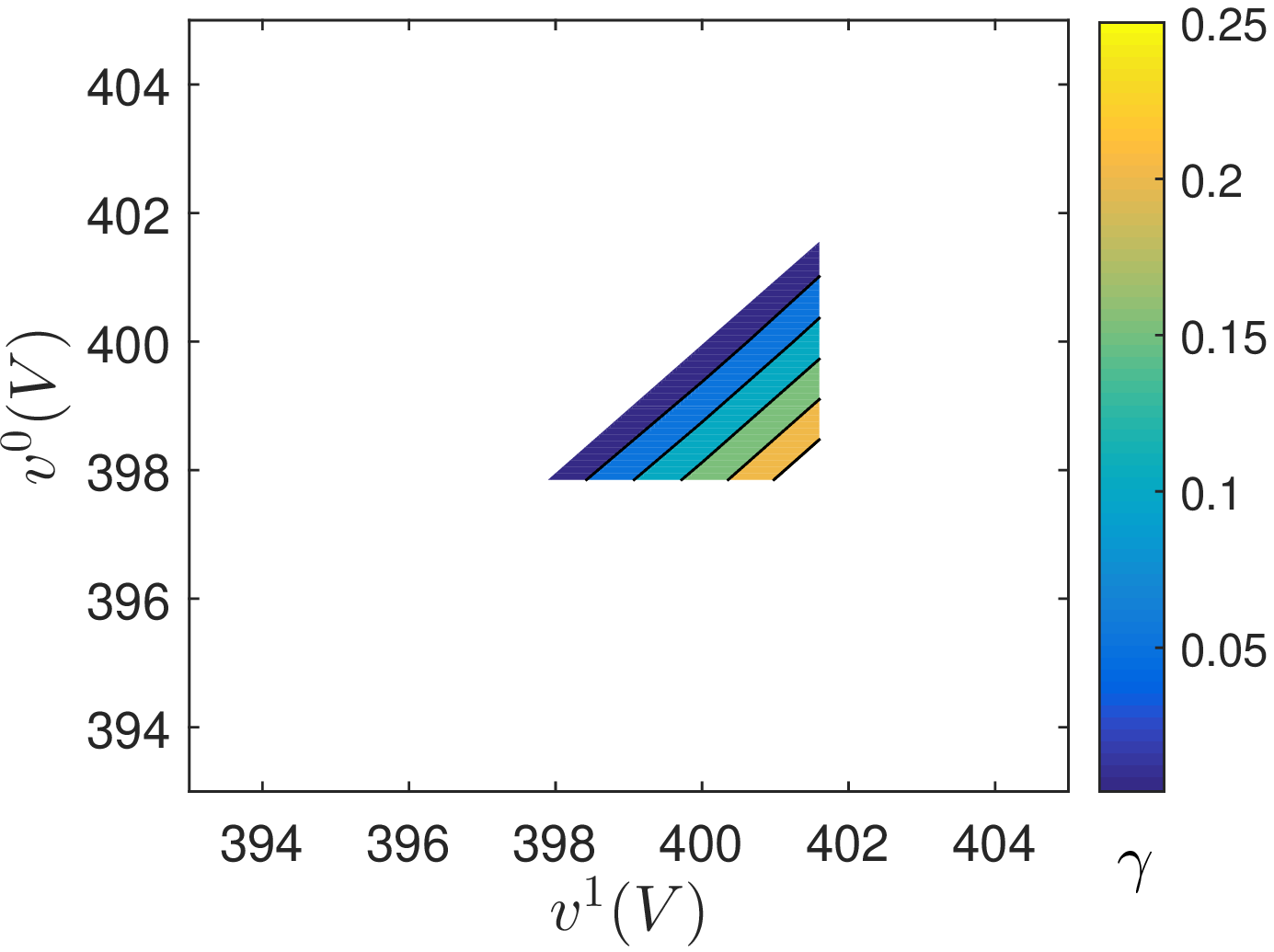}\label{SSpaceK2mw}}
\hfil
\subfloat[FD: $K=3$]{\includegraphics[scale=0.33]{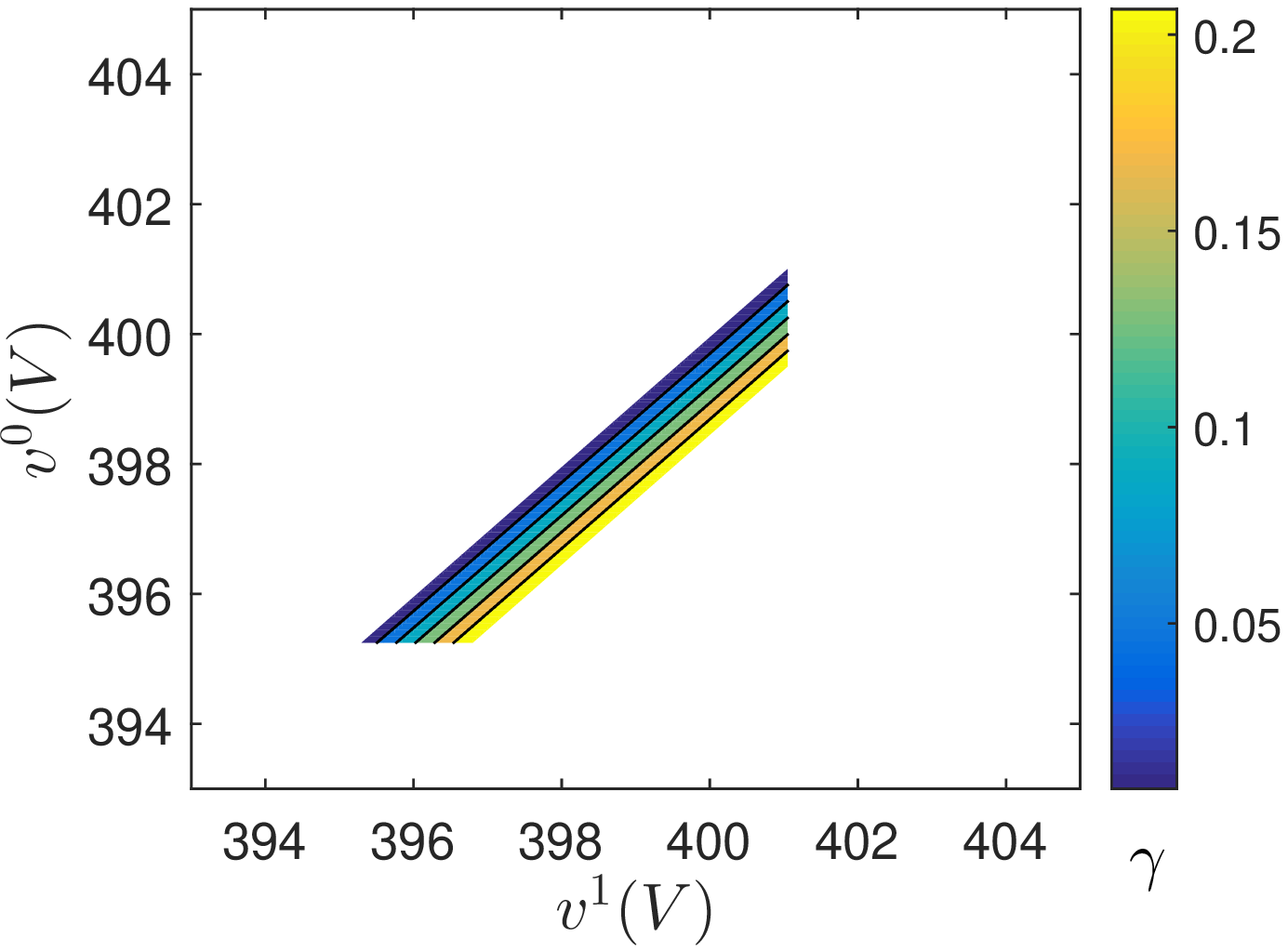}\label{SSpaceK3mw}}
\hfil
\subfloat[FD: $K=4$]{\includegraphics[scale=0.33]{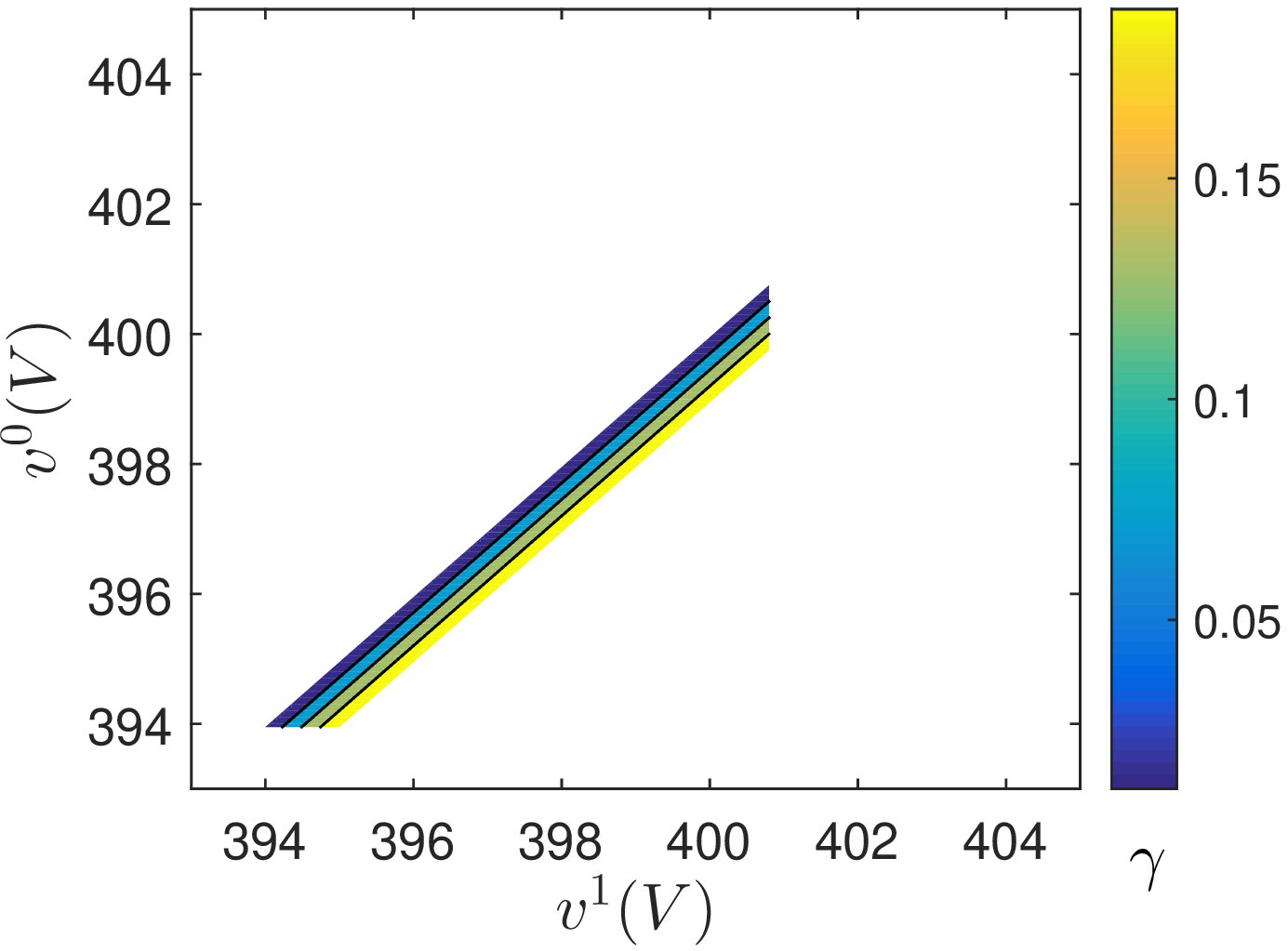}\label{SSpaceK4mw}}
\caption{{The signaling space and average power deviation. Fixed $r_{d,a}$ constellation, $r_d^1=r_d^0=2\Omega$. Equiprobable bit values.}}
\label{SSpaceK}
\end{figure*}

%We place the power talk symbols and design signaling constellations within $\mathcal{S}$, thereby guaranteeing that the operational constraints of the MG will not be violated.
Each $\mathbf{x}_k\in\mathcal{X}$ results in different output power $P_k$.
%In this mode of operation, VSC $k$ supplies power $P_{k,p}^p=v^*i_k$, since all units operate with the predefined droop parameters $\mathbf{x}_k$.
To account for this effect, we introduce the relative power deviation of unit $k$ w.r.t. its output power $P_k^{\texttt{n}}$ when \emph{all} units operate nominally: 
%when it uses $\mathbf{x}_k\in\left\{\mathbf{x}^1,\mathbf{x}^0,\mathbf{x}_k^p\right\}$ while the other units use $\mathbf{x}_i\in\left\{\mathbf{x}^1,\mathbf{x}^0,\mathbf{x}_i^p\right\},i\neq k$ w.r.t. $P_{k,p}^p$:
\begin{equation}\label{rel_pow_dev}
\delta_k(\mathbf{x}_1,\dots, \mathbf{x}_K ) = \frac{\sqrt{\mathbb{E}_{R}\left\{(P_{k}-P_{k}^{\texttt{n}})^2\right\}}}{\mathbb{E}_{R}\left\{P_{k}^{\texttt{n}}\right\}}, \; k=1,\dots,K,
\end{equation}
where $\mathbf{x}_1,\dots, \mathbf{x}_K$ are the input symbols of all units in the system (recall that output power of any unit depends on all inputs in the system), and where the averaging is performed over the load $r$, modeled as a random variable with distribution $R\sim p_R(r)$. 
%$P_{k}$ is the supplied power from VSC $k$ when it uses $\mathbf{x}_k\in\left\{\mathbf{x}^1,\mathbf{x}^0,\mathbf{x}_k^p\right\}$ while the other units use $\mathbf{x}_i\in\left\{\mathbf{x}^1,\mathbf{x}^0,\mathbf{x}_i^p\right\},i\neq k$.
The average relative power deviation of VSC $k$, and the average relative power deviation per VSC are simply:
\begin{align}
\label{delta_k}
\delta_k & = \mathbb{E}_{\mathbf{X}_1,\dots, \mathbf{X}_K} \{ \delta_k(\mathbf{x}_1,\dots, \mathbf{x}_K ) \}, \\
\delta & = \frac{1}{K} \sum_{k=1}^K \delta_k,
\end{align}
where the averaging in \eqref{delta_k} is performed over the combinations of all input symbols. %Then, the average supplied power deviation per unit $\overline{\delta}(\mathbf{x}^1,\mathbf{x}^0)$, as a function of the symbols $\mathbf{x}^0$ and $\mathbf{x}^1$ for TDMA-based power talk is calculated as
%\begin{equation}\label{dev_tdma}
%\overline{\delta}(\mathbf{x}^1,\mathbf{x}^0)=\frac{1}{2K^2}\sum_{k\in\mathcal{K}}\bigg(\sum_{\mathbf{x}_k\in\left\{\mathbf{x}^1,\mathbf{x}^0\right\}}\delta_k(\mathbf{x}_k;...,\mathbf{x}_i^p,...)+\sum_{\mathbf{x}_i\in\left\{\mathbf{x}^1,\mathbf{x}^0,\mathbf{x}_i^p\right\}}\delta_k(\mathbf{x}_k^p;...,\mathbf{x}_i,...)\bigg)
%\end{equation}
%and for MW-based power talk
%\begin{equation}\label{dev_tdma}
%\overline{\delta}(\mathbf{x}^1,\mathbf{x}^0)=\frac{1}{K2^K}\sum_{k\in\mathcal{K}}\bigg(\sum_{\mathbf{x}_k\in\left\{\mathbf{x}^1,\mathbf{x}^0\right\}}\sum_{\mathbf{x}_i\in\left\{\mathbf{x}^1,\mathbf{x}^0\right\}}\delta_k(\mathbf{x}_k;...,\mathbf{x}_i,...)\bigg)
%\end{equation}
%$\overline{\delta}(\mathbf{x}^1,\mathbf{x}^0)$ for binary power talk can be thought of as a cost assigned to each symbol pair $\mathbf{x}^1,\mathbf{x}^0$.
%The above definitions are applicable to modulations of arbitrary order.
Again, we point out that in the TDMA case, only a single input in $\mathbf{x}_1,\dots, \mathbf{x}_K$ represents an actual power talk symbol of the active unit, while the rest of them are nominal inputs.
In FD case, all inputs represent power talk symbols.
Finally, we introduce average power deviation limit $\gamma$, requiring that:
\begin{equation}\label{power_constraint}
%\overline{\delta}(\mathbf{x}^1,\mathbf{x}^0)\leq\gamma.
\delta \leq \gamma,
\end{equation}
i.e., the average power deviation per unit w.r.t. the nominal mode of operation is bound by $\gamma$.
{In TDMA case, \eqref{power_constraint} translates to an individual constraint, limiting the amount of output power deviation of each unit, whereas in FD case, it limits the average deviation of the power supplied to the load by all units jointly.}
%\PP{Here you should be more explicit about what is the constraint in the FD case vs. the TDMA case}

Fig.~\ref{SSpaceK} shows the signaling space for binary power talk, for system parameters listed in Table \ref{Param}, uniform distribution of the load $R\sim\mathcal{U}[R_{min},R_{max}]$, when both symbols are equiprobable and have fixed droop slope $r_d^1=r_d^0$ (also referred to as the fixed $r_{d}$ constellation) and $v^1>v^0$.
Evidently, TDMA power talk offers larger signaling spaces for given $\gamma$. %, which, on the other hand, provides better separation of the symbols in the detection space and more resilience to observation noise.
For both TDMA and FD, the signaling spaces decrease as the number of units increases.
%The differences in signaling spaces of TDMA and MW power talk are due to the increased number of units simultaneously communicating units in the latter.

\subsection{The detection space}\label{DSpace}

As already introduced, the detection space $\mathcal{S}_k$ for VSC $k$ is defined as the set of points $\mathbf{s}_k = (v^*,i_k)$, where $v^*$ is the output voltage, equal to the bus voltage when the line resistance is negligible, and $i_k$ is the output current of VSC $k$.
Physically, each point $\mathbf{s}_k$ represents the output power $P_k$.
By observing $\mathbf{s}_k$, VSC $k$ gathers information about the symbols/powers of other units.

\begin{figure*}[!t]
\centering
\subfloat[TDMA: $K=2$, $r=100\Omega$]{\includegraphics[scale=0.33]{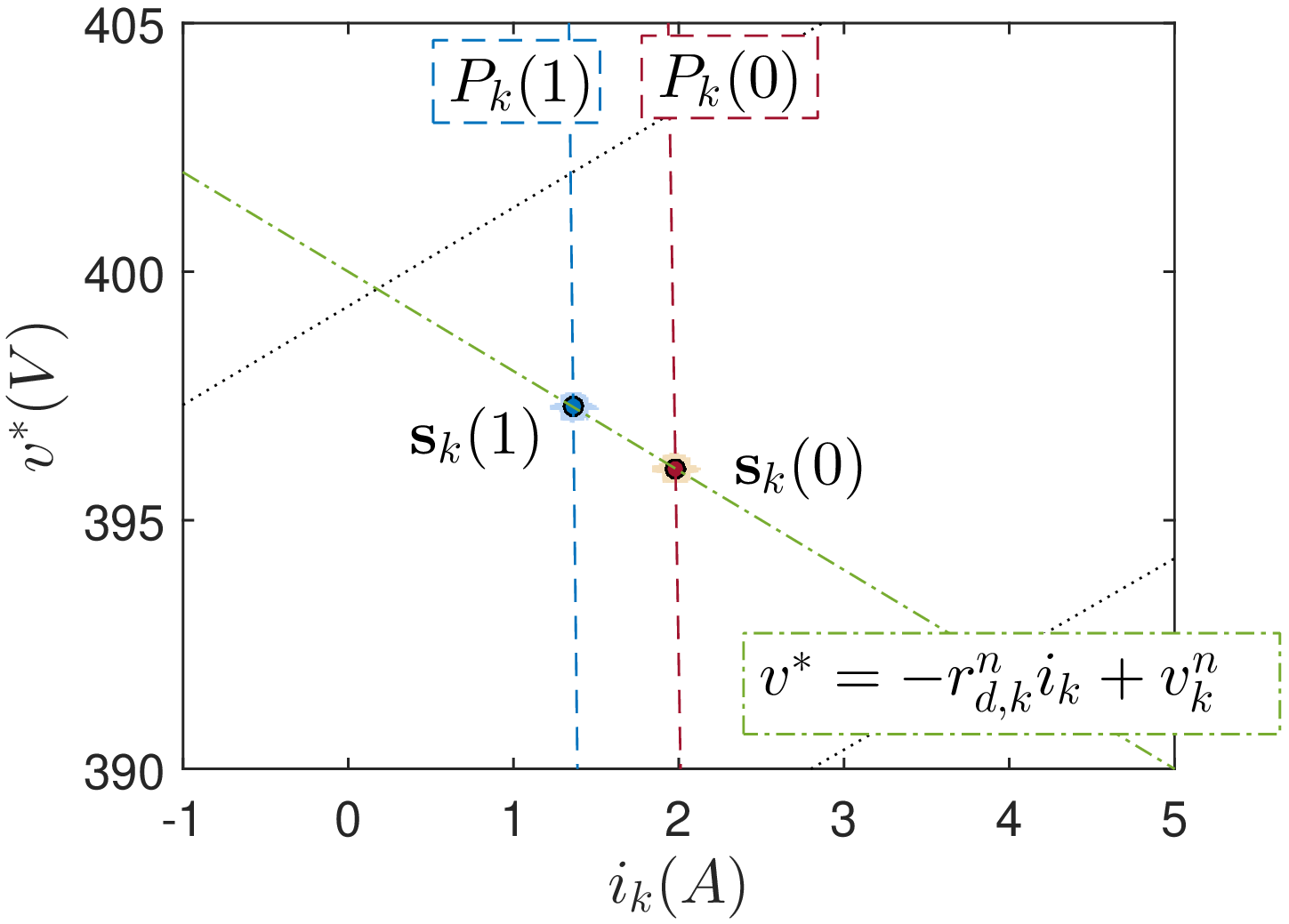}\label{DSpaceK2tdma}}
\hfil
\subfloat[TDMA: $K=2$, $r=60\Omega$]{\includegraphics[scale=0.33]{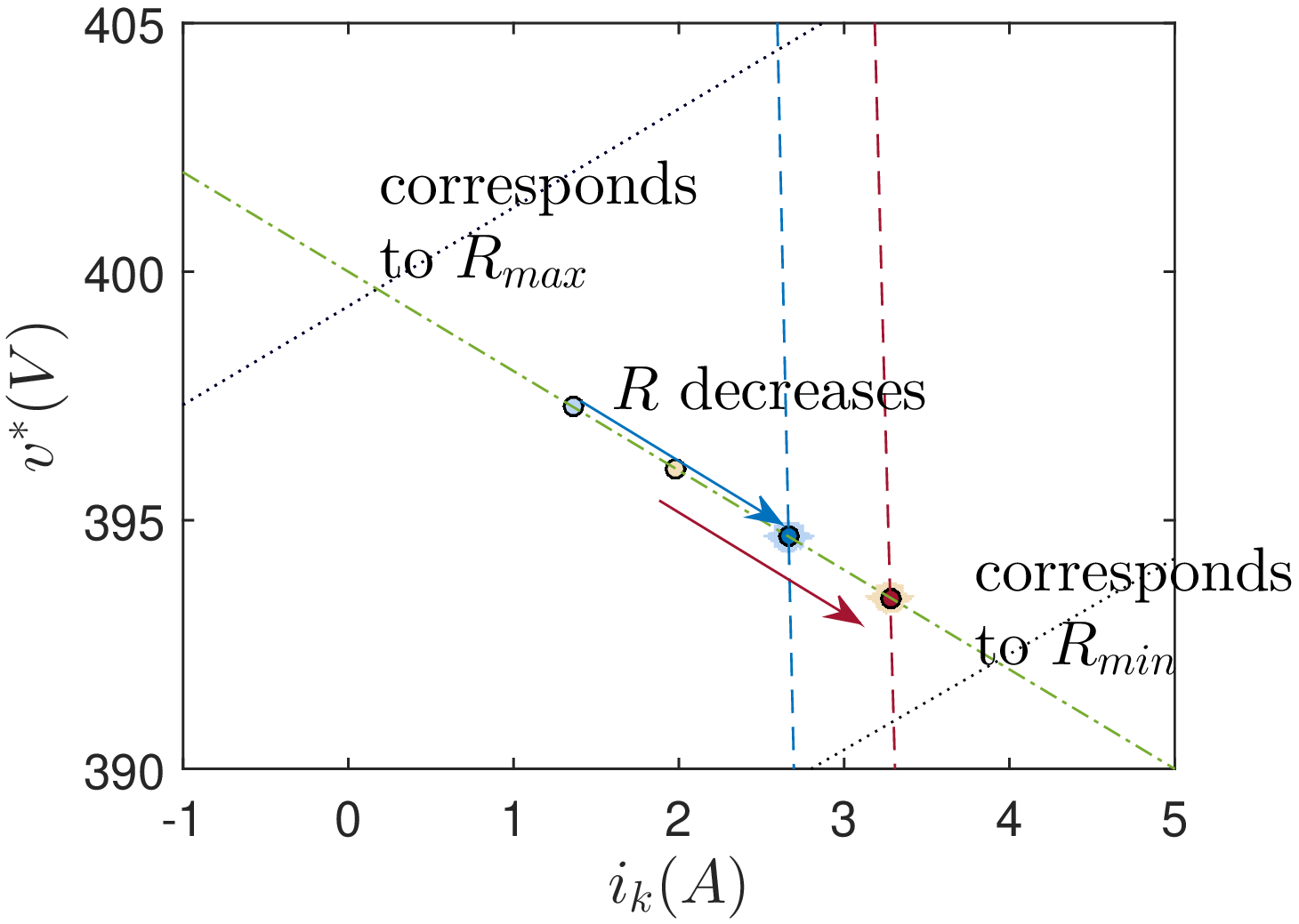}\label{DSpaceK3tdma}}
\hfil
\subfloat[TDMA: $K=4$, $r=100\Omega$]{\includegraphics[scale=0.33]{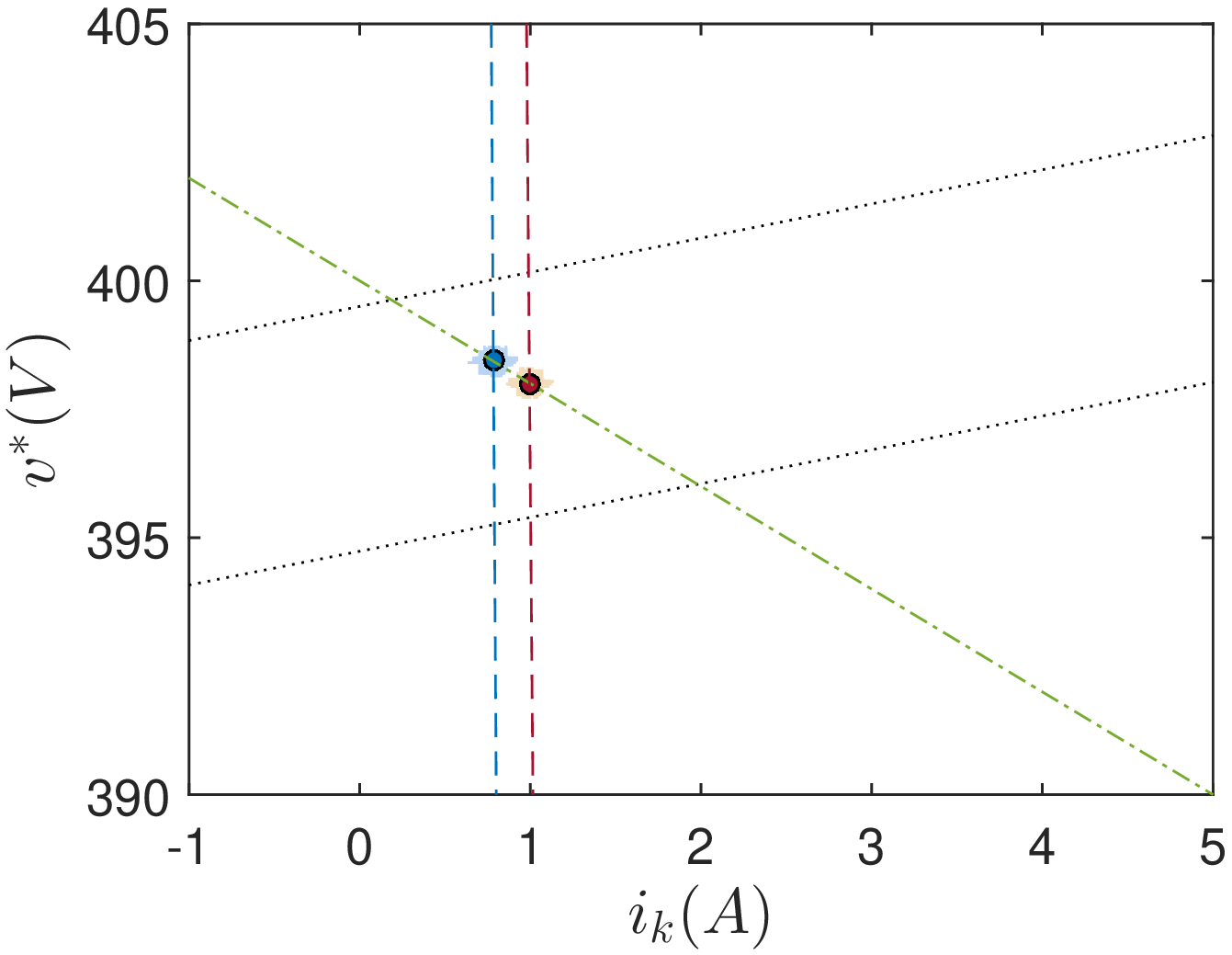}\label{DSpaceK4tdma}}
\hfil
\subfloat[FD: $K=2$, $r=100\Omega$]{\includegraphics[scale=0.33]{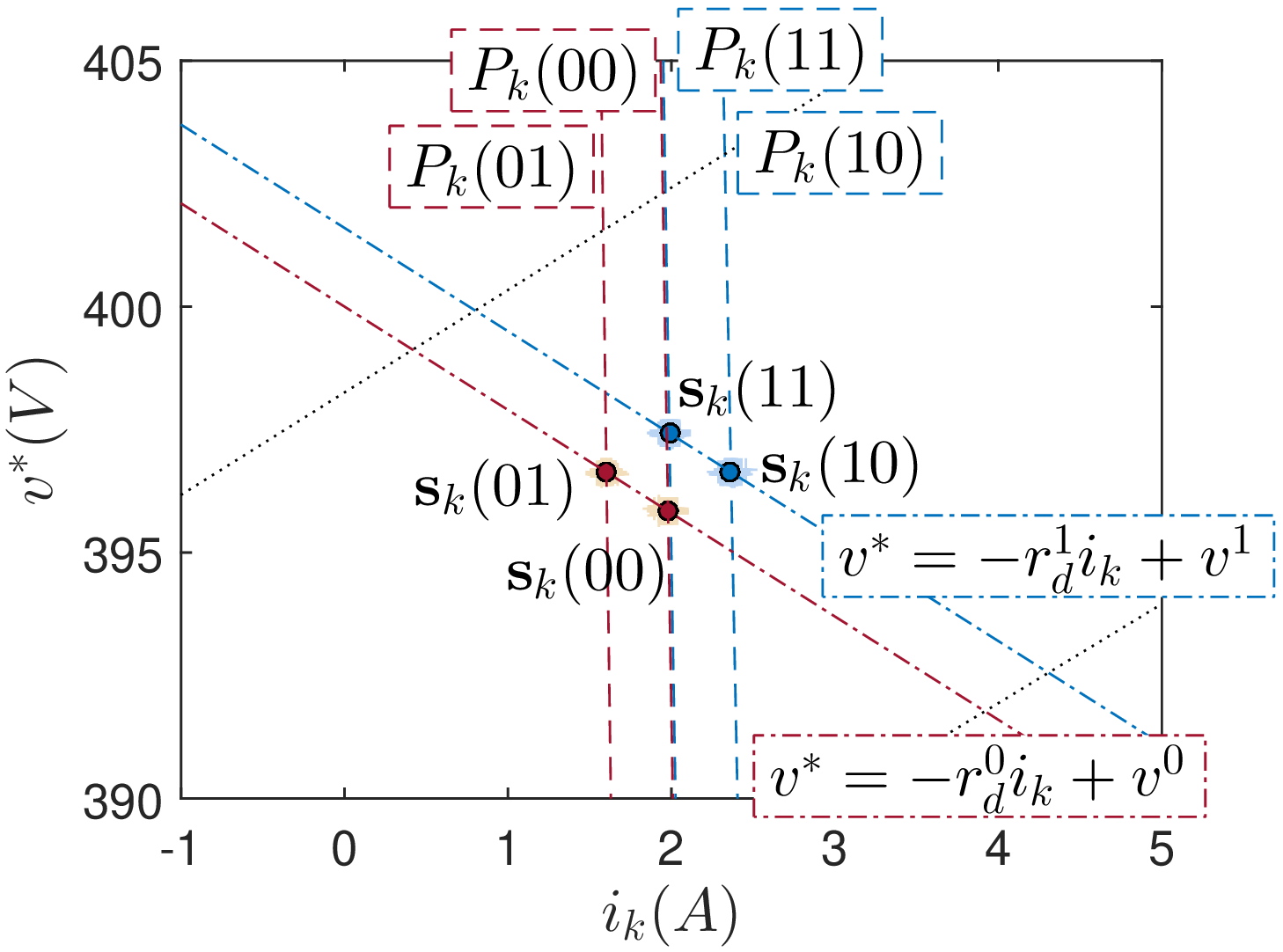}\label{DSpaceK2mw}}
\hfil
\subfloat[FD: $K=2$, $r=60\Omega$]{\includegraphics[scale=0.33]{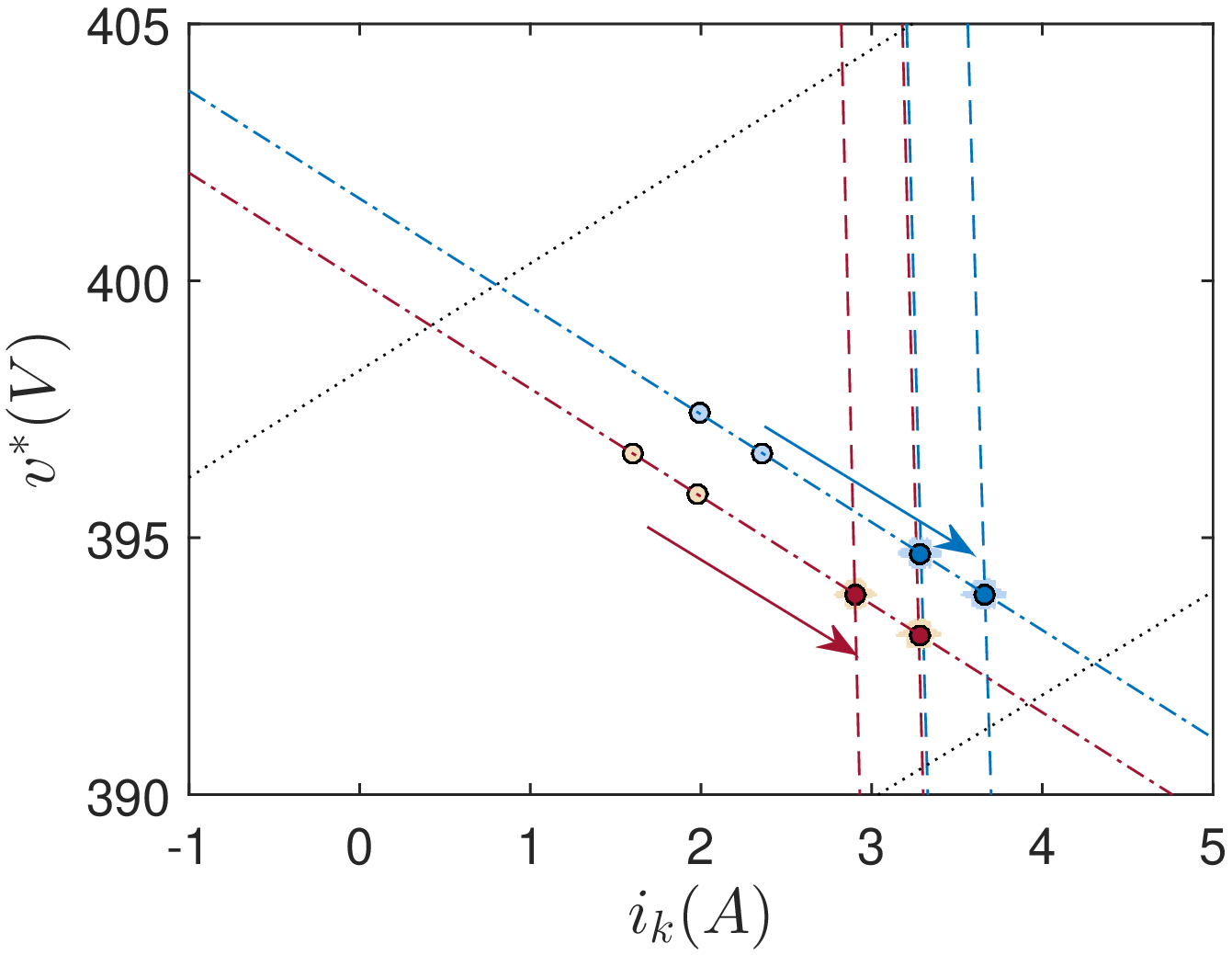}\label{DSpaceK3mw}}
\hfil
\subfloat[FD: $K=4$, $r=100\Omega$]{\includegraphics[scale=0.33]{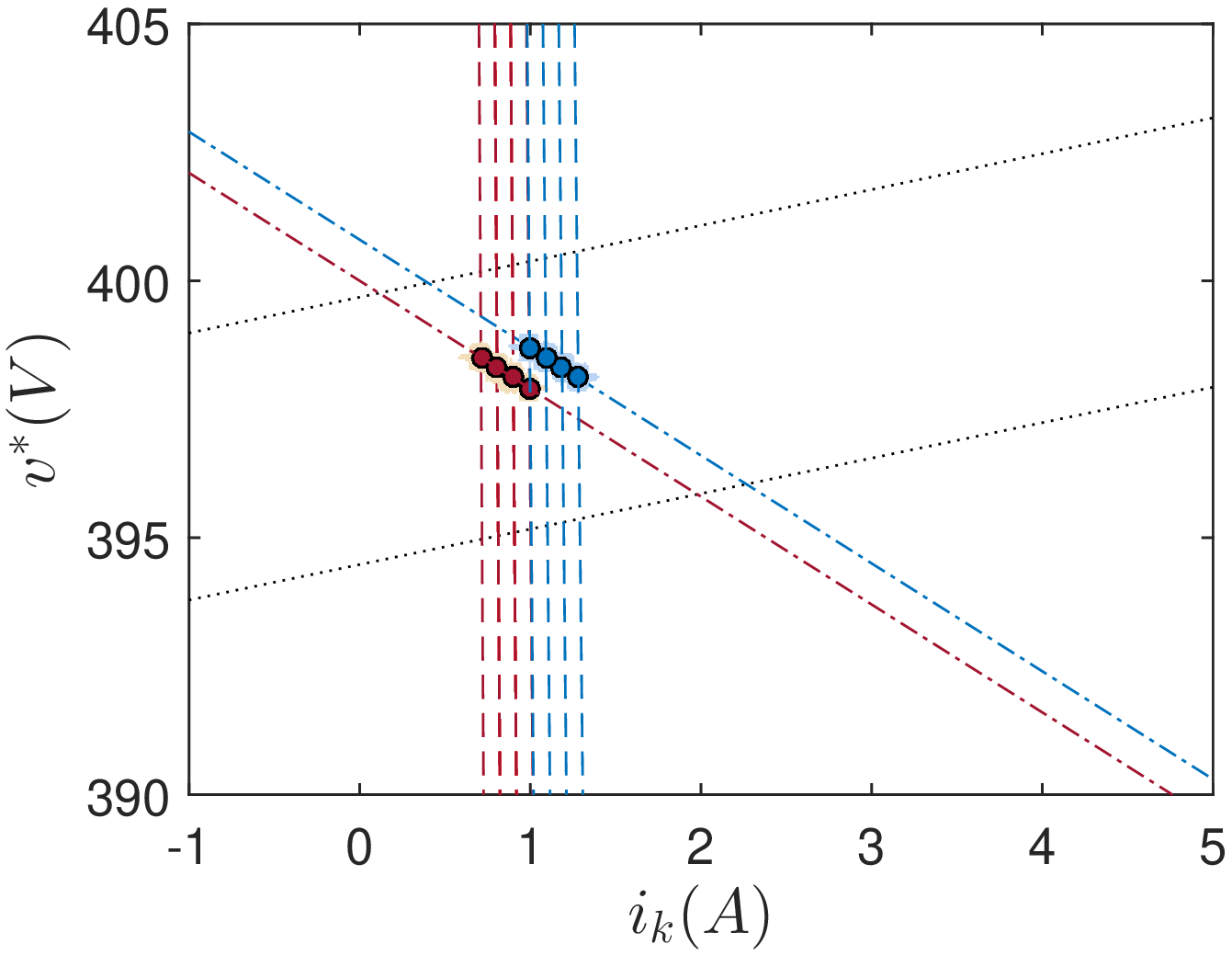}\label{DSpaceK4mw}}
\caption{The detection space of VSC $k$, fixed $r_{d}$ constellation, $\gamma=0.2$. The dashed lines represent the output power $P_k=v^*i_k$ and the dashed-dotted line $v^*=-r_{d,k}i_k+v_k$ represents the symbol $\mathbf{x}_k$ that VSC $k$ is inserting. The loci of output symbols $\mathbf{s}_k$ are on the intersection between the lines corresponding to output powers and the local inputs. As $r$ varies, $\mathbf{s}_k$ slide along the dashed-dotted lines between the bounds (dotted lines) defined by operational constraints on the load $R_{min}$ and $R_{max}$.}
\label{DSpaceK}
\end{figure*}

We start by outlining the general structure of the detection space, illustrated on Fig.~\ref{DSpaceK}. %\PP{I think a $-$ is missing on the figures for the expression $v^*=-r_{d,k}i_k+v_k$}.
All possible outputs of VSC $k$ lie on the line $v^*=-r_{d,k}i_k+v_k$ where $(v_k,r_{d,k})$ is the symbol VSC $k$ is inserting.
For TDMA power talk, if VSC $k$ is receiving, then $\mathbf{x}_k=\mathbf{x}_k^{\texttt{n}} = (v_k^{\texttt{n}},r_{d,k}^{\texttt{n}})$ and the output symbols lie on a single line, as shown in Figs.~\ref{DSpaceK}(a)-(c).
For FD power talk, a VSC sends either $\mathbf{x}^0$ or $\mathbf{x}^1$, and there are two lines on which the outputs may lie, as shown in Figs.~\ref{DSpaceK}(d)-(f).
Further, in both cases the actual loci of output symbols of VSC $k$, i.e., $\mathbf{s}_k=(v^*,i_k)$, are in the intersection of $v^*=-r_{d,k}i_k+v_k$ and $v^*=\frac{P_k}{i_k}$, where $P_k$ is the output power.
%Further, \eqref{current} reveals that $i_k$ can be reconstructed by using only the observations of the bus voltage $v^*$, which means that the information in power talk is essentially modulated in the amplitude of the bus voltage and that $i_k$ does not carry any information (refer to the discussion in Section \ref{sec:background}).
%In other words, power talk uses one-dimensional modulation in which the symbols $\mathbf{s}_k$ lie in a one-dimensional affine subspace represented with $v^*=-r_{d,k}i_k+v_k$ of $\mathbb{R}^2$.
As the load in the system varies, the points $\mathbf{s}_k$ in $\mathcal{S}_k$ shift along $v^*=-r_{d,k}i_k+v_k$, see Fig.~\ref{DSpaceK}(b) and Fig.~\ref{DSpaceK}(e).
Comparing Fig.~\ref{DSpaceK}(c) and Fig.~\ref{DSpaceK}(f) with Fig.~\ref{DSpaceK}(a) and Fig.~\ref{DSpaceK}(d), respectively, it is apparent that under constant average power deviation constraint, increasing the number of units shrinks the detection space and reduces the distance among the symbols, making the detection more susceptible to noise.
This effect is more evident for FD power talk.

Next, we describe the detection mechanism.
We assume that all VSCs, when transmitting, behave as i.i.d. Bernoulli sources, with probability of transmitting ``1'' denoted by $p_b$.
Each VSC obtains a noisy observation $\mathbf{y}_k=(\tilde{v}^*,\tilde{i}_k)$, $k = 1,\dots, K$, {see \eqref{out}}.
In TDMA binary power talk, VSC $k$ should decide between $\mathbf{s}_k(1)=(v^*(1),i_k(1))$ and $\mathbf{s}_k(0)=(v^*(0),i_k(0))$ based on $\mathbf{y}_k$, and for this purpose, employs Maximum A Posteriori Detection (MAPD) under Gaussian noise:
\begin{equation}\label{MAPD_tdma}
\ln{\frac{p(\mathbf{y}_{k}|\mathbf{s}_{k}(1),r)}{p(\mathbf{y}_{k}|\mathbf{s}_{k}(0),r)}} \underset{\mathbf{s}_{k}(0)}{ \overset{\mathbf{s}_{k}(1)} {\gtrless}} \ln{\frac{1-p_b}{p_b}}.
\end{equation}
The decision regions, i.e., the sets of points satisfying \eqref{MAPD_tdma},  are:
\begin{align}\label{MAPD_tdma_rule}
 \Lambda_{1}(r) & : \;\tilde{v}^*>\tilde{i}_{k}a_{k}^{1,0}+b_{k}^{1,0}, \text { for } \mathbf{s}_{k}(1), \\
 \Lambda_{0}(r) & : \;\tilde{v}^*<\tilde{i}_{k}a_{k}^{1,0}+b_{k}^{1,0}, \text{ for } \mathbf{s}_{k}(0),
\end{align}
where:
\begin{align}\label{a_slope}
a_{k}^{1,0} & =\frac{\sigma_{v^*}^2}{\sigma_{i_k}^2}\frac{i_{k}(1)-i_{k}(0)}{v^{*}(0)-v^{*}(1)} \\
\label{b_inter}
b_{k}^{1,0} & = \frac{1}{2}(v^{*}(0)+v^{*}(1))+\frac{1}{2}\frac{\sigma_{v^*}^2}{\sigma_{i_k}^2}\frac{(i_{k}(0))^2-(i_{k}(1))^2}{v^{*}(0)-v^{*}(1)}+\frac{\sigma_{v^*}^2}{v^{*}(1)-v^{*}(0)}\ln{\frac{1-p_b}{p_b}}
\end{align}
The error probabilities, given that the true outputs are $\mathbf{s}_{k}(1)$ and $\mathbf{s}_{k}(0)$ are, respectively:
\begin{align}\label{tdma_cross10}
\text{Pr}(e_k|\mathbf{s}_k(1),r)=\int_{\mathbf{y}_{k}\in\Lambda_{0}(r)}p(\mathbf{y}_{k}|\mathbf{s}_{k}(1),r)d\tilde{v}^*d\tilde{i}_{k}=1-Q\bigg(\frac{b_k^{1,0}-v^{*}(1)+i_k^1a_k^{1,0}}{\sqrt{\sigma_{v^*}^2+(\sigma_{i_k}a_k^{1,0})^2}}\bigg)\\ \label{tdma_cross01}
\text{Pr}(e_k|\mathbf{s}_k(0),r)=\int_{\mathbf{y}_{k}\in\Lambda_{1}(r)}p(\mathbf{y}_{k}|\mathbf{s}_{k}(0),r)d\tilde{v}^*d\tilde{i}_{k}=Q\bigg(\frac{b_k^{1,0}-v^{*}(0)+i_k^0a_k^{1,0}}{\sqrt{\sigma_{v^*}^2+(\sigma_{i_k}a_k^{1,0})^2}}\bigg)
\end{align}
Finally, the average error probability is:
\begin{equation}\label{err_n_tdma}
\text{Pr}(e_k)=\mathbb{E}_{R}\left\{\text{Pr}(e_k|\mathbf{s}_k(1),r)p_b+\text{Pr}(e_k|\mathbf{s}_k(0),r)(1-p_b)\right\}
\end{equation}

In FD binary power talk, VSC $k$ receives {$\mathbf{s}_k ( b_k, w_H )=(v^*( b_k, w_H ),i_k( b_k, w_H ))$ where $w_H( \mathbf{b}_{\sim k} )$ is the Hamming weight of the sequence of bits $\mathbf{b}_{\sim k}$ of the other units, see \eqref{FD_sk}}.
Given the local input $\mathbf{x}^{b_k}$, $\mathbf{s}_k ( b_k, w_H )$ is a priori distributed according to the Binomial distribution:
\begin{equation}\label{MW_apriori}
\text{Pr}(\mathbf{s}_k ( b_k, w_H ))={K-1 \choose w_H}p_b^{w_H}(1-p_b)^{K-1-w_H}
\end{equation} 
The MAPD decides in favor of $\mathbf{s}_k ( b_k, W_i )$ when $\mathbf{y}_{k}=(\tilde{v}^*,\tilde{i}_k)$ is observed if:
\begin{equation}\label{MAPD_mw}
\ln{\frac{p(\mathbf{y}_{k}|\mathbf{s}_k ( b_k, W_i ),r)}{p(\mathbf{y}_{k}|\mathbf{s}_k ( b_k, W_j ),r)}}\geq \ln{\frac{\text{Pr}(\mathbf{s}_k ( b_k, W_j ))}{\text{Pr}(\mathbf{s}_k ( b_k, W_i ))}},\; W_j=0,...,K-1,W_i\neq W_j
\end{equation}
The decision region for symbol $\mathbf{s}_k ( b_k, W_i )$, denoted by $\Lambda_{W_i}(r)$, is defined as:
\begin{align}\label{MAPD_mw_rule}
\Lambda_{W_i}(r) :\; \left\{
  \begin{array}{lr}
    \tilde{v}^*<\tilde{i}_{k}a_{k,b_k}^{W_i+1,W_i}+b_{k,b_k}^{W_i+1,W_i}, \\
    \tilde{v}^*>\tilde{i}_{k}a_{k,b_k}^{W_i-1,W_i}+b_{k,b_k}^{W_i-1,W_i}.
  \end{array}
\right.
\end{align}
for $0<W_i<K-1$ and:
\begin{align}\label{MAPD_me_rule1}
 \Lambda_{0}(r) & : \;\tilde{v}^*<\tilde{i}_{k}a_{k,b_k}^{1,0}+b_{k,b_k}^{1,0}, \\\label{MAPD_me_rule1}
 \Lambda_{K-1}(r) & :\;\tilde{v}^*>\tilde{i}_{k}a_{k,b_k}^{K-1,K-2}+b_{k,b_k}^{K-1,K-2} .
\end{align}
for $\mathbf{s}_k ( b_k, 0 )$ and $\mathbf{s}_k ( b_k, K-1 )$, respectively, where:
\begin{align}\label{a_slope}
a_{k,b_k}^{l,h}  = &\frac{\sigma_{v^*}^2}{\sigma_{i_k}^2}\frac{i_{k}(b_k,l)-i_{k}(b_k,h)}{v^{*}(b_k,h)-v^{*}(b_k,l)} \\
\label{b_inter}
b_{k,b_k}^{l,h} = & \frac{1}{2}(v^{*}(b_k,h)+v^{*}(b_k,l))+\frac{1}{2}\frac{\sigma_{v^*}^2}{\sigma_{i_k}^2}\frac{(i_{k}(b_k,h))^2-(i_{k}(b_k,l))^2}{v^{*}(b_k,h)-v^{*}(b_k,l)} +  \nonumber \\
& + \frac{\sigma_{v^*}^2}{v^{*}(b_k,h)-v^{*}(b_k,l)}\ln{\frac{\text{Pr}(\mathbf{s}_{k}(b_k,l))}{\text{Pr}(\mathbf{s}_{k}(b_k,h))}}
\end{align}
and $l,h=0,...,K-1,l\neq h$. 
The probability of error when the true output is $\mathbf{s}_k ( b_k, W_i )$ is:
\begin{equation}\label{err_n}
\text{Pr}(e_k|\mathbf{s}_k ( b_k, W_i ),r) = \int_{\mathbf{y}_{k}\notin\Lambda_{W_i}(r)}p(\mathbf{y}_{k}|\mathbf{s}_k ( b_k, W_i ),r)d\tilde{v}^*d\tilde{i}_{k}
\end{equation}
and the average error probability can be calculated as:
\begin{equation}\label{err_n_mw}
\text{Pr}(e_k)=\mathbb{E}_{R}\left\{\sum_{\mathbf{x}^{b_k}\in\left\{\mathbf{x}^1,\mathbf{x}^0\right\}}\sum_{W_i=0}^{K-1}\text{Pr}(e_k|\mathbf{s}_k ( b_k, W_i ),r){K-1 \choose W_i}p_b^{W_i}(1-p_b)^{K-1-W_i}\text{Pr}(\mathbf{x}^{b_k})\right\}
\end{equation}

\begin{figure}[t]
\centering
\includegraphics[scale=0.55]{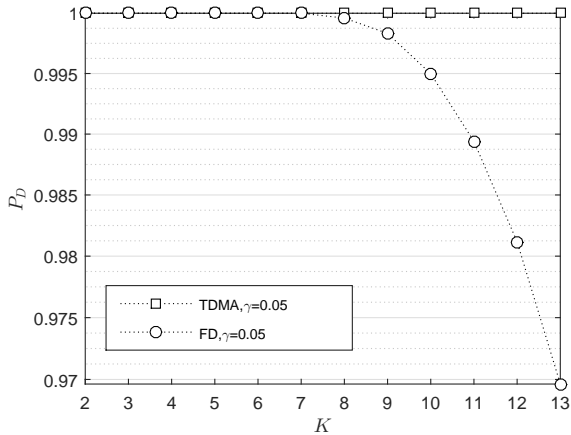}
\caption{{Probability of correct detection. Fixed $r_d$ constellation, $r_d^1=r_d^0=2\Omega$, $v^0=400V$, $v^1$ chosen to satisfy $\delta=\gamma$, equiprobable bit values.}}
\label{Noise}
\end{figure}

As already noted, the observations $\mathbf{y}_k$, $k=1,\dots,K$, are obtained as averages over a slot.
The variance of the observation noise $\sigma^2 = \text{diag}\left\{\sigma_{v^*}^2,\sigma_{i_k}^2\right\}$, {see Section~\ref{GeneralComm}}, depends on the slot duration $T_s$ and the sampling frequency $f_o$.
As $T_s$ is of the order of milliseconds and $f_s$ of the order of kHz, $\sigma_{v^*}^2$ and current $\sigma_{i_k}^2$ have rather modest values.
%\textcolor{blue}{Nevertheless, as shown in Figs.~\ref{DSpaceK}(a) and (f) , the effective detection space and distances between symbols decrease as the number of units increase, and the observation noise may have impact on the detection reliability.
Fig.~\ref{Noise} illustrates the probability of correct detection $P_D = 1- \frac{1}{K} \sum_{k=1}^{K} P(e_k)$ as the number of units in the system $K$ increases.
The values $\sigma_{v^*}=\sigma_{i_k}=0.001$ corresponding to the worst case scenario of the observation noise variance for the measurement equipment used in modern low voltage MGs \cite{ref:19,ref:20,ref:21,ref:22,ref:23}, when $T_s=1 \, \text{ms}$ and $f_o = 10 \, \text{kHz}$.
The symbols are chosen from the fixed $r_{d}$ constellation such that the average relative power deviation $\gamma=0.05$ is satisfied, i.e., the constellation is \emph{not} specifically optimized to minimize the error probability.
Obviously, the proposed detector performs exceptionally well in case of TDMA power talk.
In FD version of the scheme, the detection is also practically errorless for $K \leq 8$.
We also note that for larger $\gamma$ the effects of noise become practically negligible both for TDMA and FD binary power talk.
Taking into account the above results, it is evident that the need for the additional mechanisms to combat observation noise is rather modest, if the proposed detection method is employed.

We conclude by noting that the decision regions are established for a given value of the load $r$, and they need to be reset when $r$ changes.
This could done using training sequences, which are sent by VSCs in a coordinated manner, and which enable construction of $\mathcal{S}_k$, $k = 1, \dots, K$.
Details on this aspect of power talk are given in the next section.

\section{Designing Power Talk Communication Protocols}\label{Protocols}

%We have already illustrated the effect of the changes of the load, i.e. the system state on the detection space and we have pointed out that a feasible strategy to deal with such changes is to reconstruct the detection after state change using sequence of training bits that will reset the new points $\mathbf{s}_k$ in the detection space of VSC $k$.
In this section, we present techniques that complement the basic power talk operation, as described in the previous Sections, required to design a  fully operational communication protocol.
To evaluate the efficiency of the proposed solutions, we use the net transmission rate per unit $\eta$, i.e., average number of information bits transmitted per unit per time slot. %and (ii) the net reception rate per unit $\mu$, i.e., average number of information bits received per unit per slot.
In further text, we assume a MG with $K$ units and fair scheduling.
%\begin{align}\label{eff_gen}
%\eta = \frac{1}{K}\sum_{k=1}^{K}\eta_k, \; \; \mu=\frac{1}{K}\sum_{k=1}^{K}\sum_{i\neq k} \eta_i, 
%\end{align}
%where $\eta_k$ is the net transmission rate of unit $k$.
%The above net rates take into account all the overheads related to protocol operation.

\subsection{Coding for multiple access}\label{Code}

Here we compare the rates of TDMA and FD versions of power talk, denoted by $\eta^\texttt{S}_{\text{TDMA}}$ and $\eta^\texttt{S}_{\text{FD}}$, respectively when $r$ is stable, i.e., it does not change, and the detection spaces are accurate.
In TDMA power talk, the transmissions are orthogonal in time and the net transmission rate can be simply written as:
\begin{align}\label{eta_tdma}
\eta^\texttt{S}_{\text{TDMA}} = 1 / K.
\end{align}
%We denote the number of redundant bits necessary to transmit a single bit in multiple access settings with $C(K)-1$.
%Normally, it depends on $K$.
%With TDMA, the transmissions are orthogonal in time and there is no need to code for MAI, however each bit transmitted by each unit in the system is effectively associated with $K-1$ redundant bits stemming from the lost transmission opportunities, when other units in the system are transmitting.
%Thus, $C_{TDMA}(K)=K$.
%
In FD power talk, we deal with BI-MAAC, where a unit observes sums of the bits transmitted by $K-1$ units.
{Chang and Weldon in \cite{ref:28}} proposed a reference coding solution for this type of multiple access, which enables unique decodability of user codewords and asymptotically achieves the maximum sum rate of the BI-MAAC.
%Each unit has a codebook with only two codewords, one for each bit.
Table \ref{table1} lists the achievable transmission rates per unit per slot $\eta^S_{\text{FD}}$ for the scheme from \cite{ref:28}, again assuming stable operation. %for unique decodibility at the receiver as a function of the number of units.
%We denote this table with $CW$.
Evidently, from Table~\ref{table1} it follows: 
\begin{align}
\eta_{\text{FD}}^\texttt{S} & >  \eta_{\text{TDMA}}^\texttt{S} = 1 / K.
\end{align}
%Moreover, the reception rate per unit is the sum of the transmission rates of the remaining units:
%\begin{align}\label{mu_mw}
%\mu_{\text{FD}} = & ( K - 1 ) \eta_{\text{FD}} > ( K - 1 ) \mu_{\text{TDMA}}. 
%\end{align}
Therefore, FD power talk is more efficient in terms of resource utilization, when the effects related to re-establishment of the detection space when the changes of the load $r$ are neglected.
%However, a trade-off arises in power talk as a result of the necessity to learn and reset the detection space as the state of system changes.
%The trade-off captures the sum rate the corresponding scheme provides and the cost that needs to be paid to communicate with this sum rate, expressed though the general metric $\overline{\eta}$ defined with \eqref{eff_gen}.

\begin{table}
\caption{Achievable rates of Chang-Weldon codes \cite{ref:28} for BI-MAAC}
\label{table1}
\centering
\begin{tabular}{|c||c|c|c|c|c|c|c|c|c|c|c|c|c|c|c|c|c|c|}
\hline
$K$ & 1 & 2 & 3 & 4 & 5 & 6 & 7 & 8 & 9 & 10 & 11 & 12 & 13 & 14 & 15 & 16 & 17 & 18 \\ \hline
$\eta^S_{\text{FD}}$ & 1 & 1/2 & 1/2 & 1/3 & 1/3 & 1/4 & 1/4 & 1/4 & 1/5 & 1/5  & 1/6  & 1/6  & 1/6  & 1/7  & 1/7  & 1/8  & 1/8  & 1/8  \\ \hline
\end{tabular}
\end{table}

\subsection{Training phase}\label{TrainCode}

%Both protocols are investigated in TDMA and MW settings as described above.
%Since in this section we design communication protocols to deal with variable system state and shifting detection space as the state changes (subsection \ref{DSpace}).
%The training of the system to construct the detection space locally at each VSC unit can be realized using different strategies.
All VSCs have to maintain a layout of the detection space that matches the current value of the load $r$.
A simple and effective strategy for the construction of the detection space is to conduct a training phase during which the units input predefined training  sequences.
We assume that each unit builds its detection space separately, in order to take into account imperfections of the MG system, such as small resistances of the feeder lines and the common bus.%, that stem from the heterogeneity of the system.

%Simple and robust strategy is to use TDMA approach, i.e. each unit in the system learns the systems and constructs its detection space using $a$ dedicated time slots.
%$a$ should be number larger than $1$ since the detection space should be constructed with high fidelity and reliability.
Denote the length of the training phase in slots by $L$. 
%We have already announced the differences between TDMA and MW realization of the power talk in terms of necessary overhead to construct the detection space locally at each unit (subsections \ref{Basic2},\ref{DSpace}).
In TDMA binary power talk, a unit has to learn 2 points in its detection space for each of the remaining units, see Section~\ref{Basic}.
%Thus, each VSC unit in principle has to discover $2$ points in its own detection space for each transmitting unit.
Assuming that a point is learned during $M$ dedicated slots,\footnote{In general, using $M > 1$ improves the reliability of the detection space construction, at a cost of an increased training phase.} the  training phase length is $L_{\text{TDMA}}= 2  M K$ slots.
The training phase can be performed by sequential transmission of $\mathbf{x}^0$ and $\mathbf{x}^1$ by one of the units, while the remaining units operate nominally and construct their detection spaces.

In FD-based power talk, the number of outputs each VSC can observe in the detection space is $2K$, see Section~\ref{Basic}.
%\textcolor{red}{In MW power talk, the training phase should be performed such that units simultaneously transmit bits that make up sequences of characteristic Hamming weights and simultaneously build-up their detection spaces.}
Then, the total number of points to be learned is $2K^2$, and, assuming that a point is learned during $M$ slots, the training phase length is $L_{\text{FD}}=2MK^2$.
The training can be conducted such that a units construct their detection spaces sequentially, through coordinated transmissions with other units.
Evidently, the FD binary power talk requires $K$ times more slots than the TDMA version.\footnote{For homogenous and close to ideal systems, all units can construct their detection spaces simultaneously, making the training phase significantly shorter.
It can be shown that the minimum number of slots necessary to build all detection spaces in an ideal single bus MG, assuming $M$ slots per point, is $4M$ for TDMA and $4KM$ for FD power talk.}% Still, MW variant of the scheme needs at $K$ times more slots.}

We now turn to potential schemes for activation of the training phase.
A simple solution is to perform training periodically and update the detection spaces.
A more involved approach would be to employ a model change detector that tracks the bus voltage and, if a change is detected, initiates the training phase.
In further text, we investigate and compare the above two schemes.
For this purpose, we model the load changes through a Poisson process whose intensity $\lambda$ is the expected number of load changes per time slot.
%Thus, the probability of load change occurrence during a single slot of duration $T_s$ is $p=1-e^{-\lambda}$.
As the slot duration $T_s$ is of the order of milliseconds, one can expect that $\lambda<<1$ in practice.

\subsubsection*{Power Talk with periodic training phase}\label{Periodic}

\begin{figure}[t]
\centering
\includegraphics[scale=0.6]{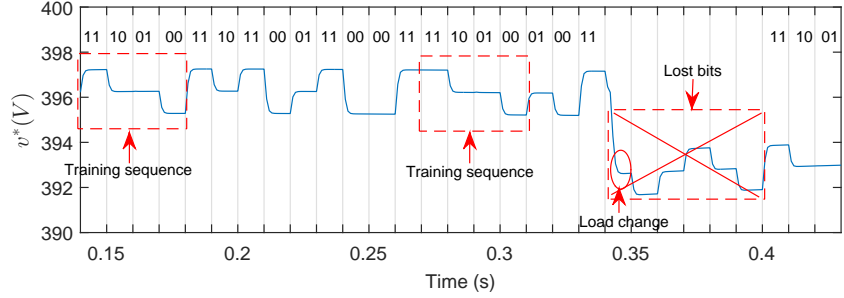}
\caption{Power talk with periodic training. $T_s = 1 \, \text{ms}$.}
\label{ProtI}
\end{figure}
\begin{figure}[t]
\centering
\includegraphics[scale=0.6]{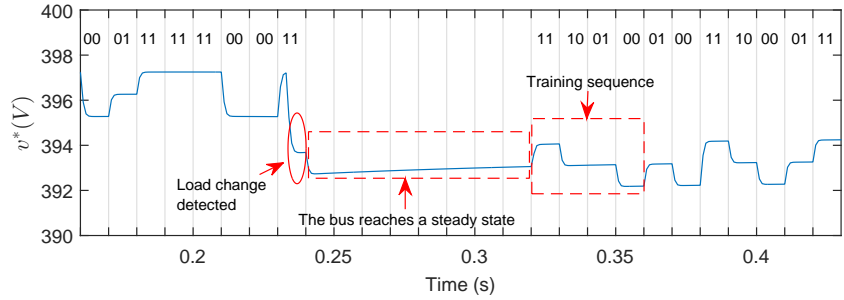}
\caption{Power talk with with load change detection. $T_s = 1 \, \text{ms}$.}
\label{ProtII}
\end{figure}

We assume that the training phase occurs periodically after each VSC transmits $B$ bits of information.
Also, we assume the worst-case scenario in which each system state change completely ``destroys'' all detection spaces and all following transmissions are lost, see Fig.~\ref{ProtI}.
%In particular, if the state changes during the training phase, all the following transmissions are lost, and if the state changes after the training phase during the data phase, the bits after the change are considered to be lost.
%Denote the actual number of bits sent by each VSC (bits that are correctly received by all other units) $B_k^{'}$.
%With these considerations, definition \eqref{eff_gen} can be rewritten as follows:
%\begin{equation}\label{eff_pI}
%\overline{\eta}=\frac{1}{K}\sum_{k=1}^K\frac{\overline{B^{'}_k}}{L+\frac{B}{\eta_b}}(1-p)^{L}.
%\end{equation}
%$\overline{B^{'}}_k$ is the average number of bits sent by unit $k$ in one training+data cycle with duration $L+\frac{B}{\eta_b}$.
%Note that $\frac{B}{\eta_b}$ gives the number of slots, necessary to transmit $B$ bits in multiple access communication settings.

In Appendix~\ref{pITDMA} it is shown that the net transmission rate for TDMA power talk is:
\begin{equation}\label{eff_pI_tdma}
\eta_{\text{TDMA}}=\frac{(1-p)^{L_{\text{TDMA}}+1}[1-(1-p)^{KB}]}{p(L_{\text{TDMA}}+KB)} \eta^\texttt{S}_{\text{TDMA}},
\end{equation}
where $p = 1 - e^{-\lambda}$ is the probability that the load changes during a slot and $\eta^\texttt{S}_{\text{TDMA}}$ is the corresponding stable rate, {see \eqref{eta_tdma}}.
For FD variant, the net transmission rate is:
\begin{equation}\label{eff_pI_mw}
\eta_{\text{FD}} = \frac{(1-p)^{L_{\text{FD}}+\frac{1}{\eta^\texttt{S}_{\text{FD}}}}[1-(1-p)^{\frac{B}{\eta^\texttt{S}_{\text{FD}}}}]}{[1-(1-p)^{\frac{1}{\eta^\texttt{S}_{\text{FD}}}}](L_{\text{FD}}+\frac{B}{\eta^\texttt{S}_{\text{FD}}})},
\end{equation}
where $\eta^\texttt{S}_{\text{FD}}$ is the stable transmission rate, see Table \ref{table1}.
When $p\rightarrow 0$, $\eta \rightarrow \frac{B}{L \eta^\texttt{S} +  B} \eta^\texttt{S}$, both for TDMA and FD variants.
Also, from \eqref{eff_pI_tdma} and \eqref{eff_pI_mw}, one can find the optimal length of the bit block $B$ for given $\lambda$ and $K$, where $B_{opt}=\max_{B} \eta(B;K,p)$.
%This, in general leads to intractable equations that are difficult to solve in closed form.
%$\overline{\mu}$ can be simply written as $\overline{\mu}=(K-1)\overline{\eta}$ since the net rates $\overline{\eta}_k$ are equal.
%Finally, net reception rates can be found similarly as in \textcolor{red}{\eqref{eta_tdma} and \eqref{mu_mw}}.

\subsubsection*{Power Talk with load change tracker}\label{Detector}

Each VSC can, in principle, locally track $v^*$ and $i_k$ and decide whether the system state has changed; the state changes should be detected by all units simultaneously with high reliability.
An option is to use a standard model change detector \cite{ref:26} that tracks the voltage level.
Assume that each VSC implements such state change detector, operated it in the following way: i) if a change is detected, then the current transmission is stopped, %inserts the training sequence, if the  after successful resetting of the detection space, the transmission is resumed, 
ii) $L_{\text{BS}}$ {``blank slots''} (e.g., nominal operation symbols $\mathbf{x}_k^{\texttt{n}}$) are inserted by all units, to allow the system to reach steady state after the load change and, iii) the training phase is re-initiated; as illustrated in Fig.~\ref{ProtII}.
Appendix \ref{appB} derives the following expressions for net transmission rate:
\begin{align}\label{eff_pII_tdma}
{\eta}=\frac{1}{p+(1-p)^{-(L+L_{\text{BS}})}} \eta^\texttt{S},
\end{align}
which holds both for TDMA and FD power talk.
%Evidently, $\overline{\eta}$ does not depend on the length of the data block $B$.
When $p\rightarrow 0$, $\eta \rightarrow \eta^S$.
%$\overline{\mu}$ can be simply written as $\overline{\mu}=(K-1)\overline{\eta}$.

\section{Performance Evaluation}\label{perf}

%This section evaluates the proposed power talk.
%We ignore the effect of the noise, and illustrate the performance of the two protocols, introduced in Section \ref{Protocols} in terms of dealing with random load variations for TDMA and MW realizations.
The performance evaluation is obtained using the MG system described in Section II.
The bits are equiprobable, i.e., $p_b=0.5$ and we use $\gamma=0.1$ to design the input symbol constellation; the effects of the observation noise in this case are virtually negligible, {see Section~\ref{Lim}}.
%However, since in the analysis in Section \ref{Protocols} we ignore the effect of the noise, the obtained results are valid for any value of $\gamma$.

\subsection{Dealing with load variations}

\begin{figure*}[!t]
\centering
\subfloat[$K=10.$]{\includegraphics[scale=0.55]{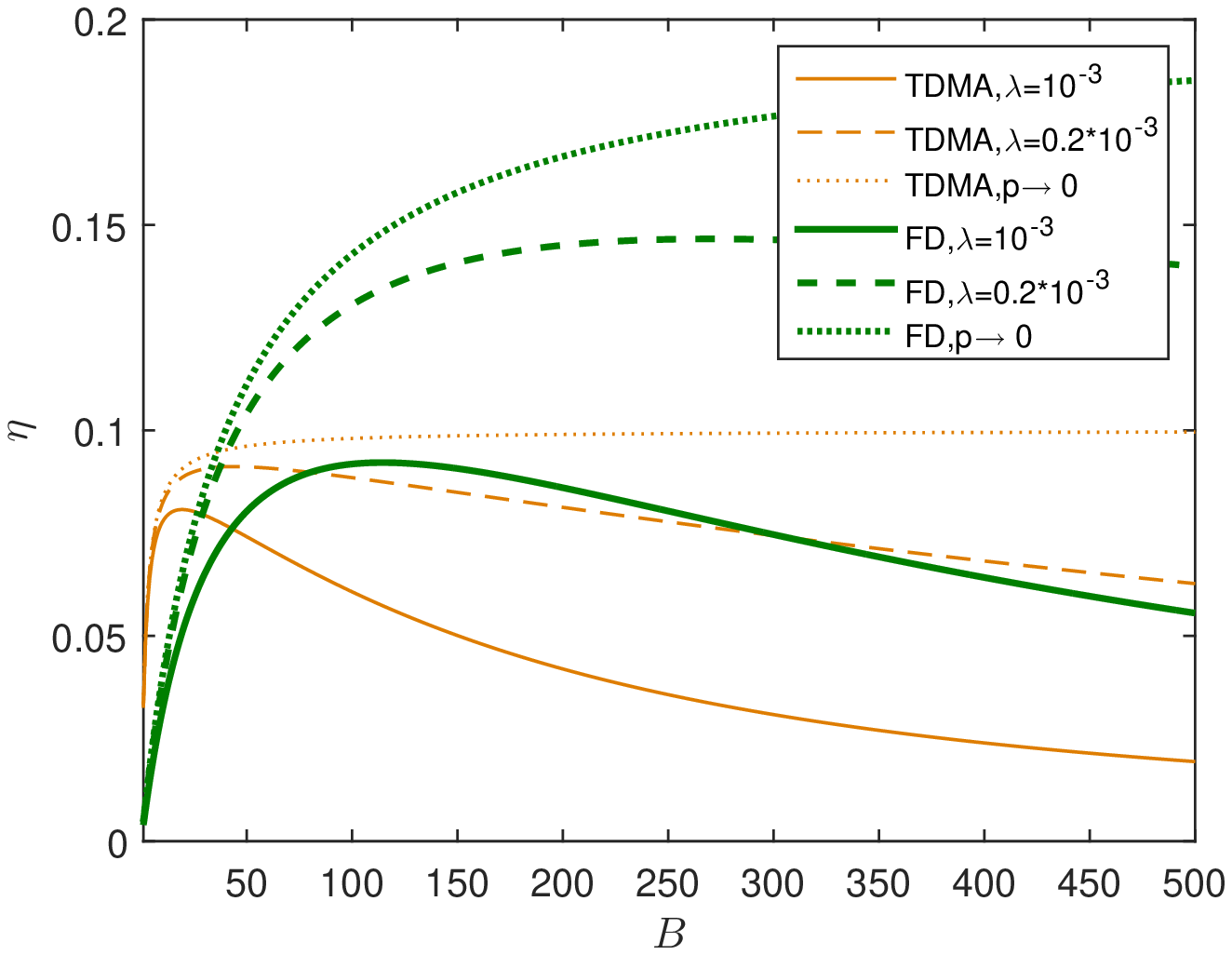}\label{eta_PITDMA}}
\hfil
\subfloat[$K=15.$]{\includegraphics[scale=0.55]{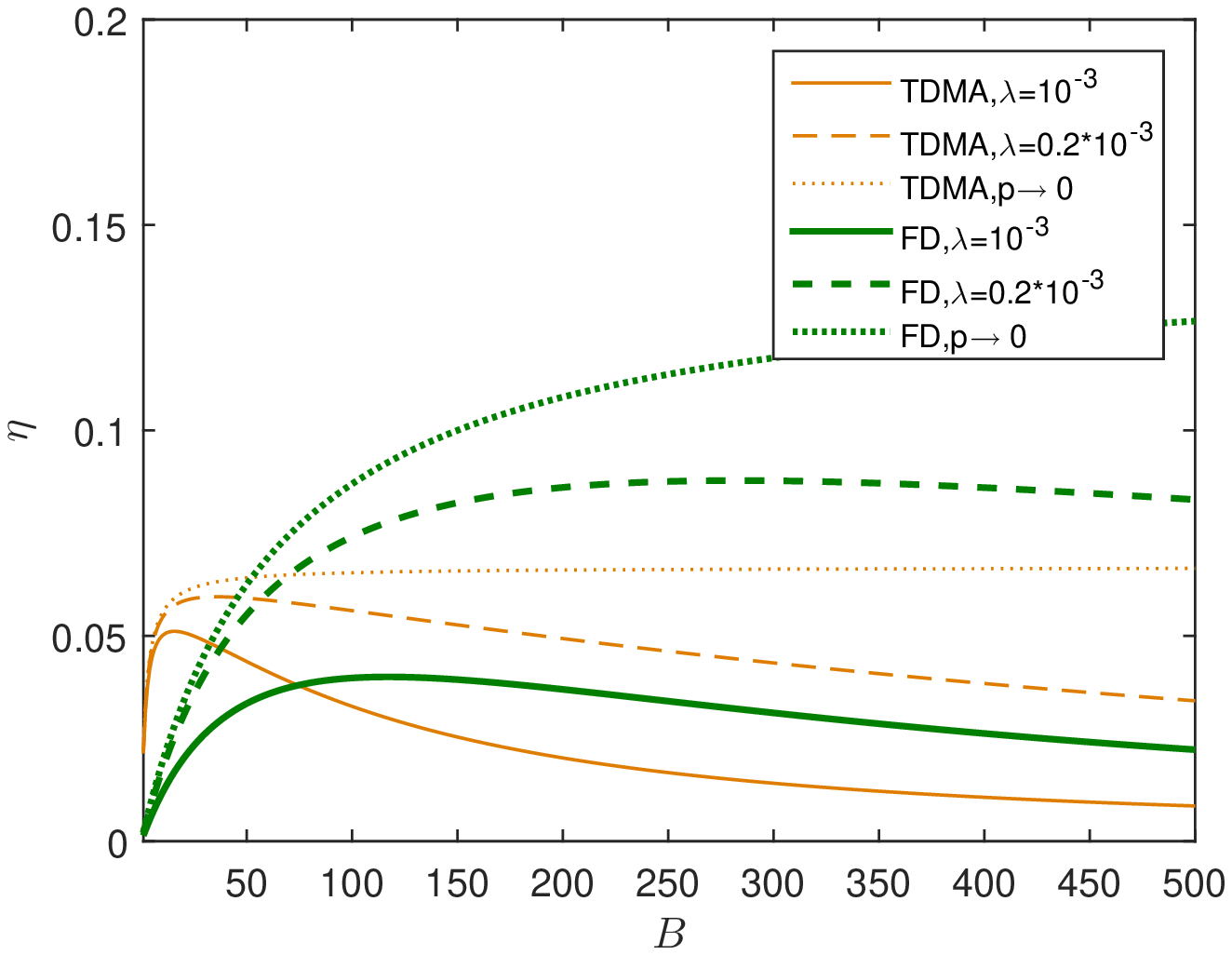}\label{eta_PIMW}}
\caption{$\eta$ as a function of $B$, $K$ and $\lambda$: periodic training sequence insertion.}
\label{eta_PI}
\end{figure*}

\begin{figure*}[!t]
\centering
\subfloat[TDMA power talk.]{\includegraphics[scale=0.55]{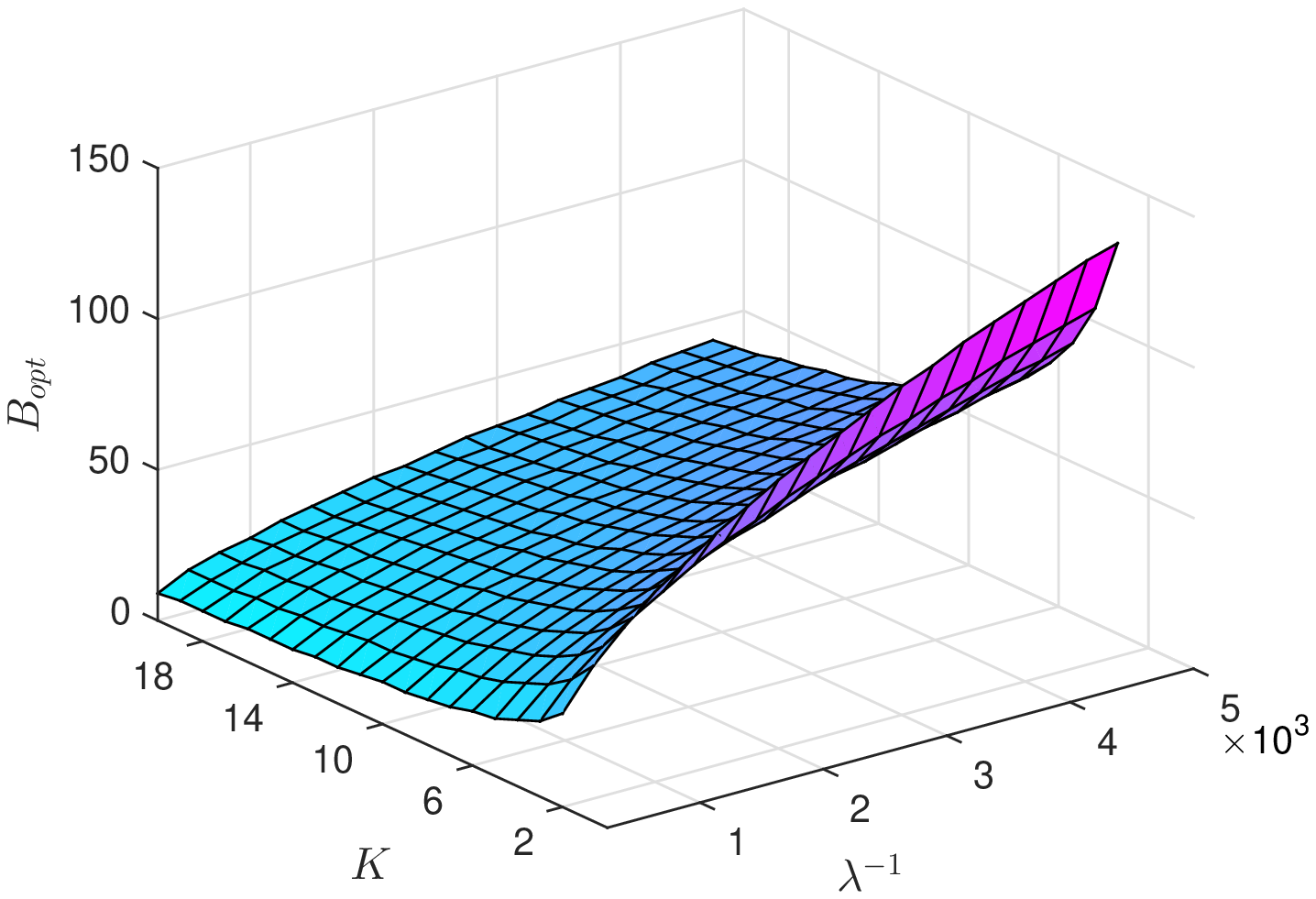}\label{BoptTDMA}}
\hfil
\subfloat[FD power talk.]{\includegraphics[scale=0.55]{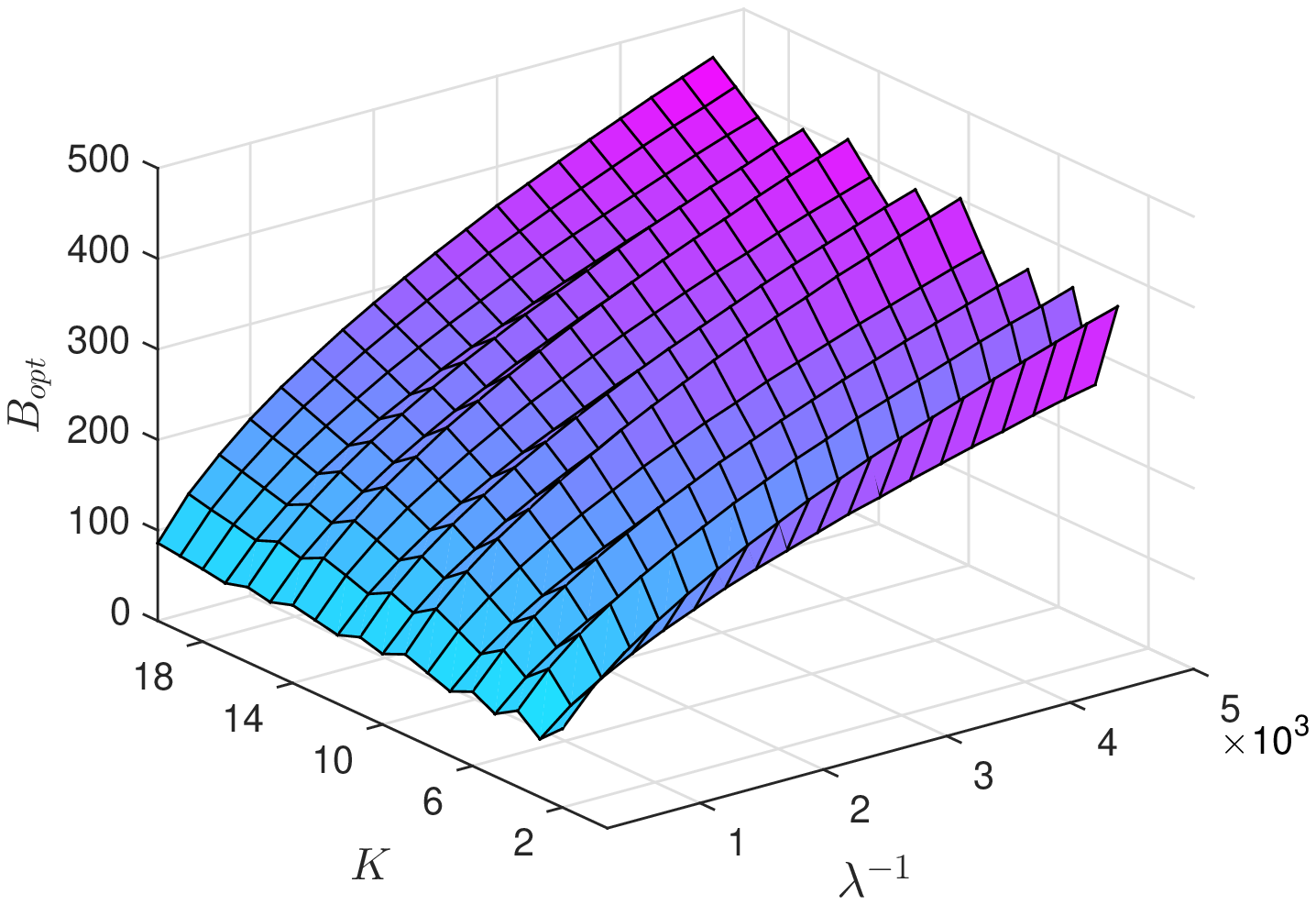}\label{BoptMW}}
\caption{The optimal choice of $B$.}
\label{Bopt}
\end{figure*}

Fig.~\ref{eta_PI} illustrates the net transmission rate per unit $\eta$ for power talk with periodic training phase, as function of $B$ which is the number of transmitted bits after the training phase is re-initiated.
In general, for small $B$, the TDMA power talk outperforms the FD variant, and this is more pronounced when the number of units $K$ in the system increases.\footnote{Note that, although the shown net transmission rates seem small, one should bear in mind that in the system with fair scheduling, a maximum transmission rate of $1/K$ can be achieved using TDMA approach, as illustrated in the figure.}

The optimal choice of $B$ that maximizes $\eta$, as function of $K$ and $\lambda$, is depicted in Fig.~\ref{Bopt}.
We note that the depicted values of $B$ are obtained using numerical evaluation; also, the fluctuations of $B_{opt}$ in case of FD variant are due to the behavior of the optimal coding rate which is a step function of $K$, {see Table~\ref{table1}}.
%An optimal number of bits that maximizes the average number of bits received per unit slot, $B_{opt}=\max_{B}\overline{\eta}(B;K,p)$ exists for both TDMA- and MW-based power talk with periodic insertion of the training sequence.
%The resulting equation is transcendent and obtaining closed form expression for $B_{opt}$ is difficult.
Clearly, FD power talk implementation requires longer $B$ to maximize the rate, as it needs to compensate for the overhead of the training phase, i.e.,  $L_{\text{FD}}>L_{\text{TDMA}}$.
Thus, FD power talk is more suitable for longer data sequences when periodic training phase is used.

\begin{figure*}[!t]
\centering
\subfloat[$K=10.$]{\includegraphics[scale=0.55]{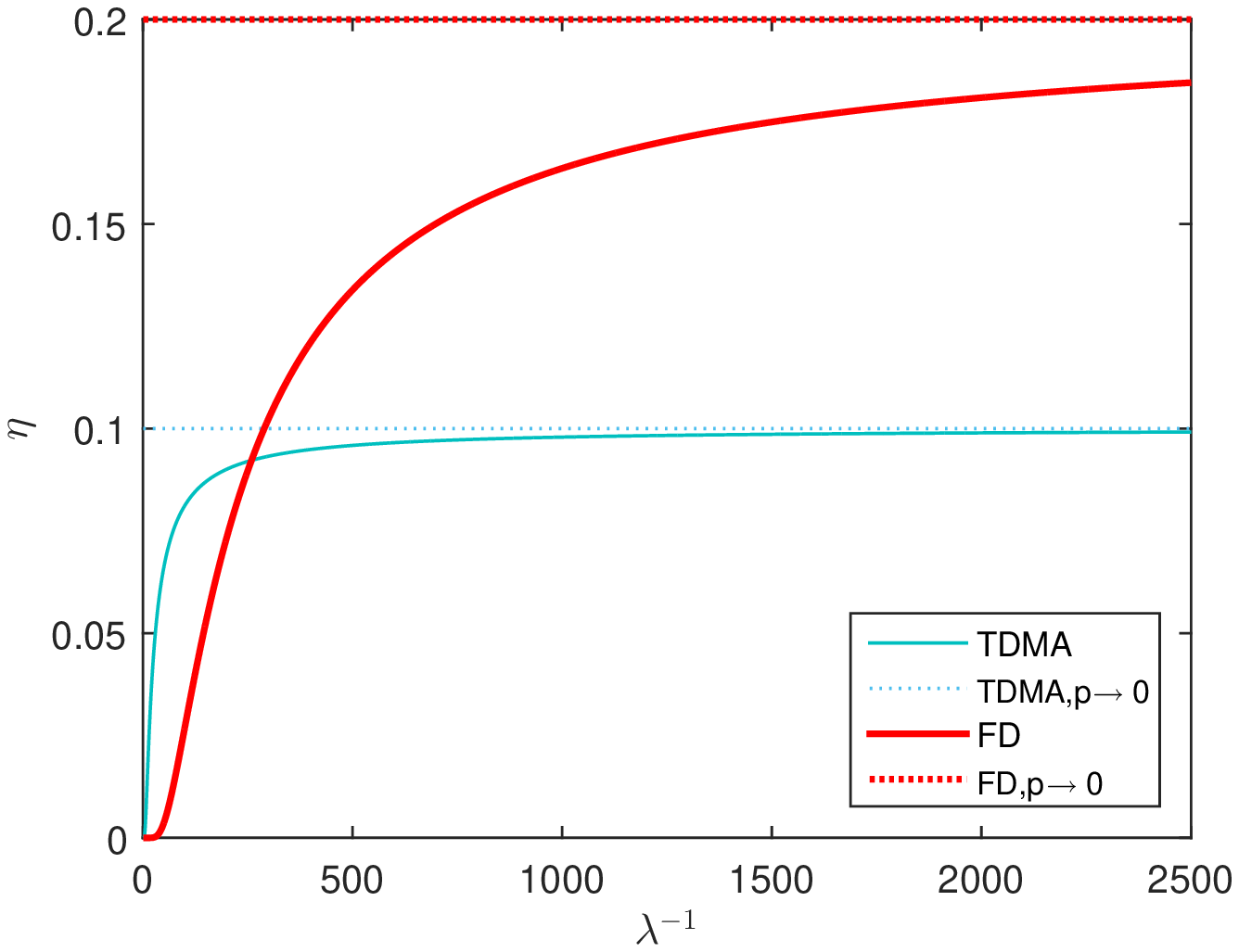}\label{eta_PIMW}}
\hfil
\subfloat[$K=15.$]{\includegraphics[scale=0.55]{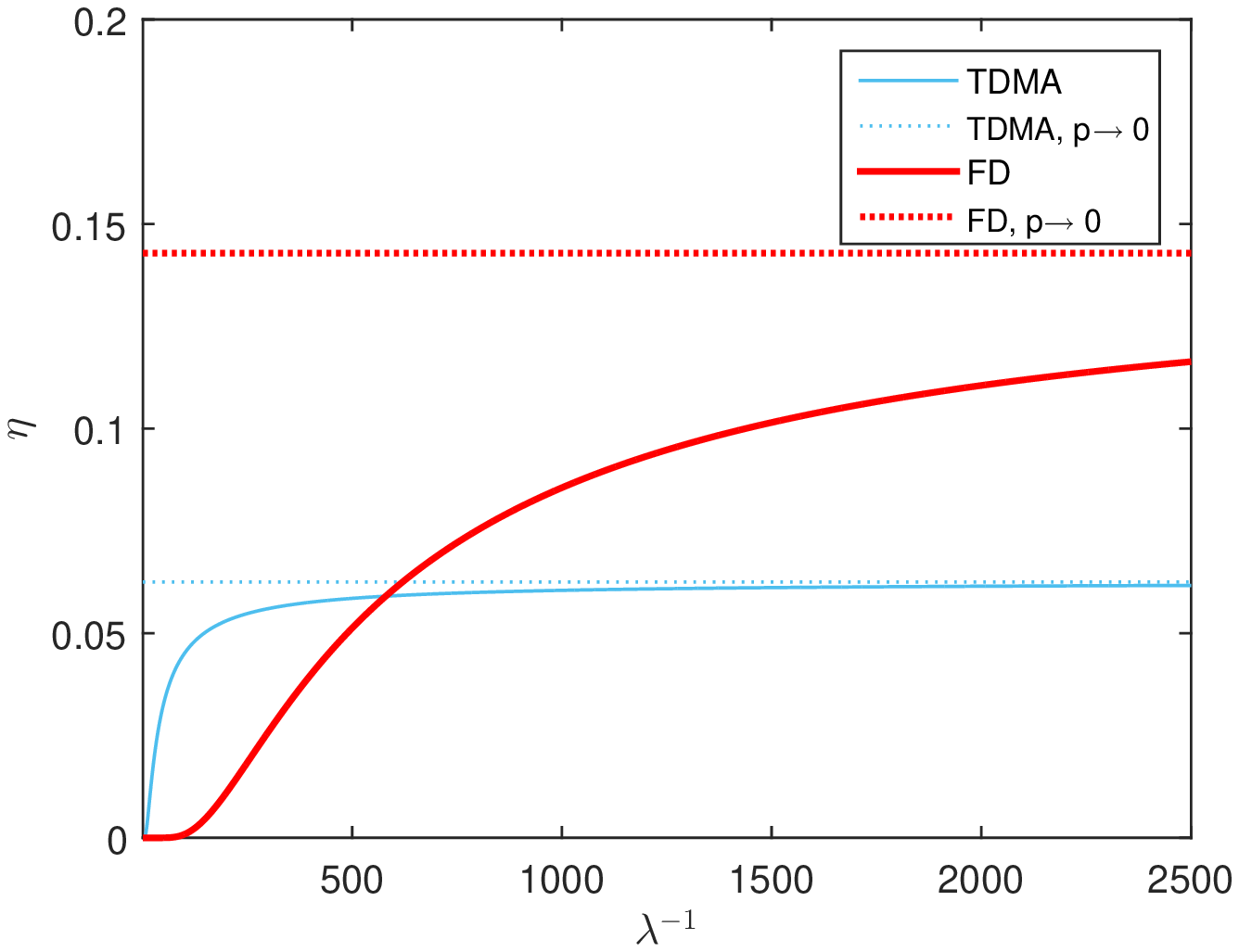}\label{eta_PITDMA}}
\caption{$\overline{\eta}$ as a function of $\lambda$: power talk with load change tracker.}
\label{eta_PII}
\end{figure*}

Fig.~\ref{eta_PII} illustrates the net transmission rate per unit $\eta$ for power talk with load change tracker as function of $\lambda$.
Evidently, the FD power talk outperforms the TDMA variant for load change intensities of practical interest, i.e., small $\lambda$.
However, TDMA power talk achieves the asymptotic performance faster, which is also a consequence of the length of the training phases.

\begin{figure*}[!t]
\centering
\subfloat[Periodic training phase.]{\includegraphics[scale=0.55]{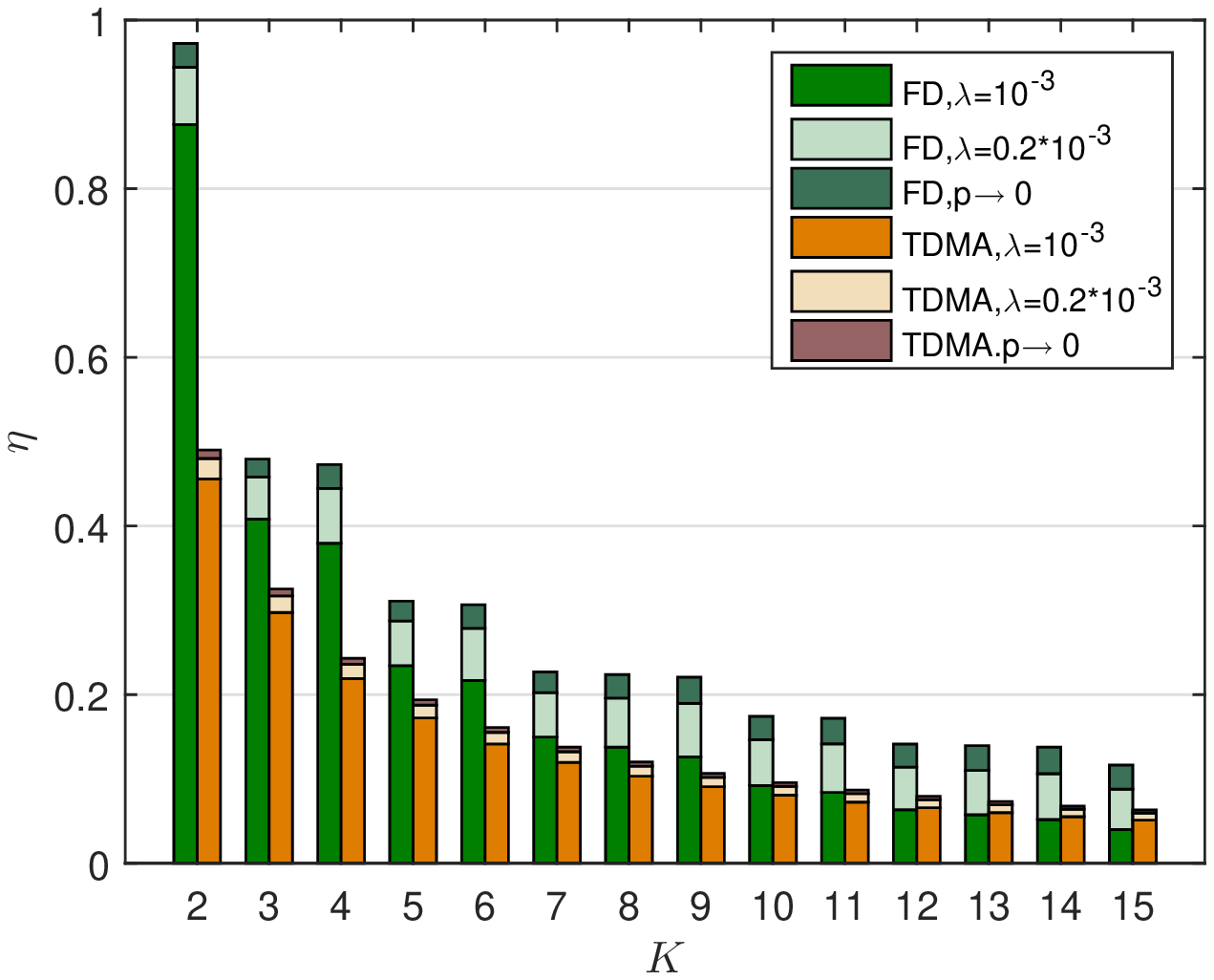}\label{etaVSK_pI}}
\hfil
\subfloat[Load change tracker.]{\includegraphics[scale=0.55]{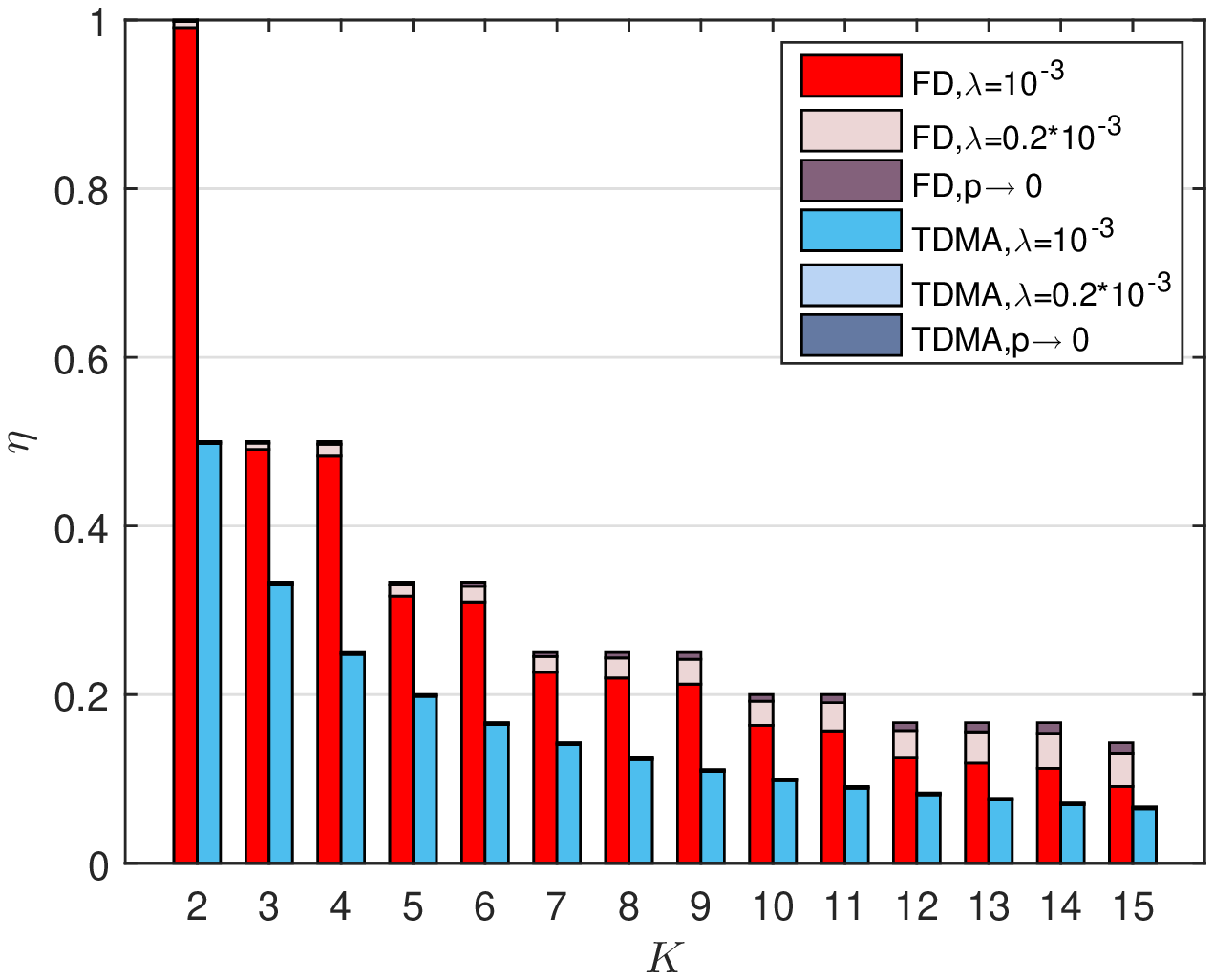}\label{etaVSK_pII}}
\caption{FD vs TDMA power talk for increasing $K$: net transmission rate per unit $\eta$.}
\label{etaVSK}
\end{figure*}
\begin{figure*}[t]
\centering
\subfloat[Periodic training phase.]{\includegraphics[scale=0.55]{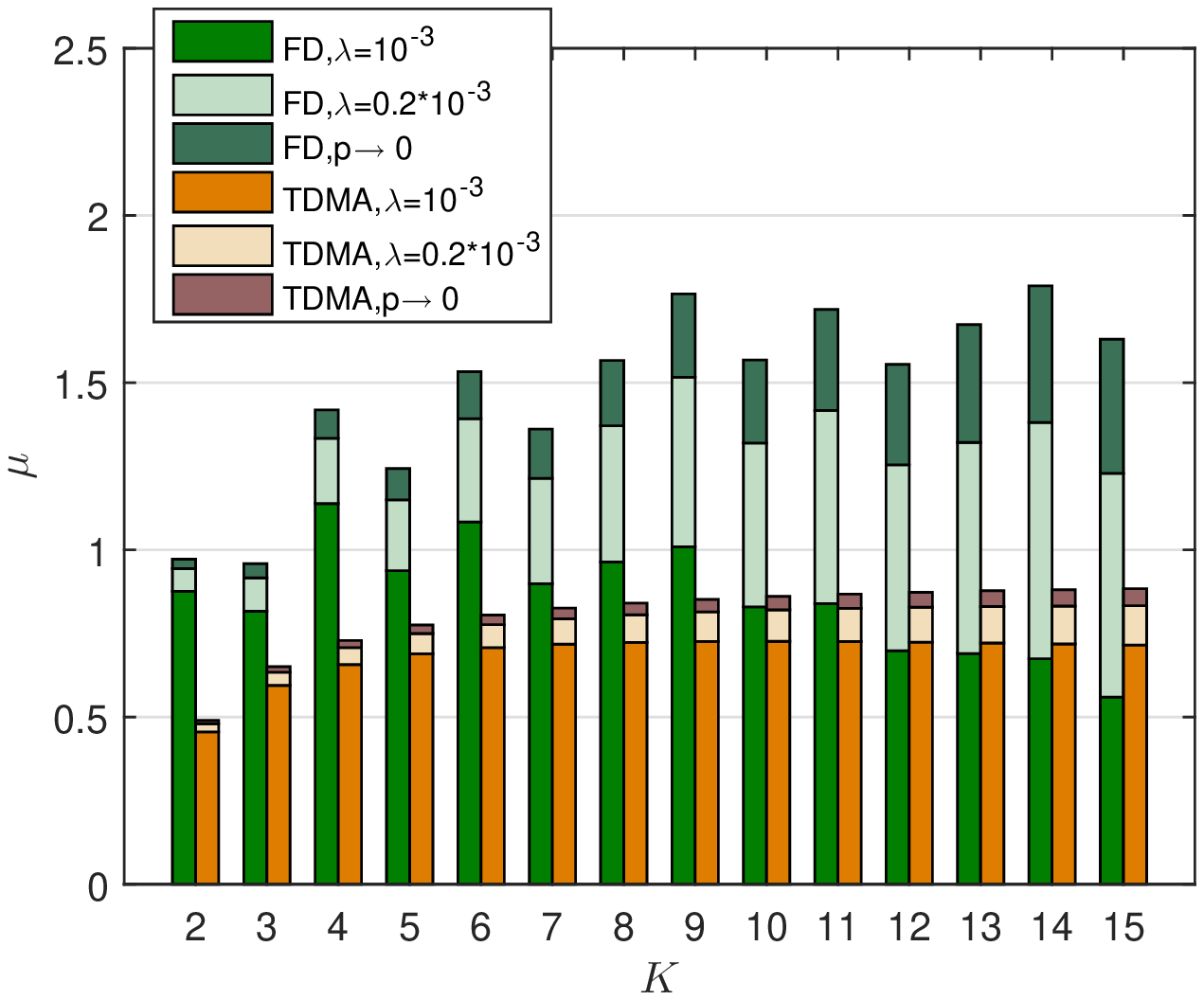}\label{muVSK_pI}}
\hfil
\subfloat[Load change tracker.]{\includegraphics[scale=0.55]{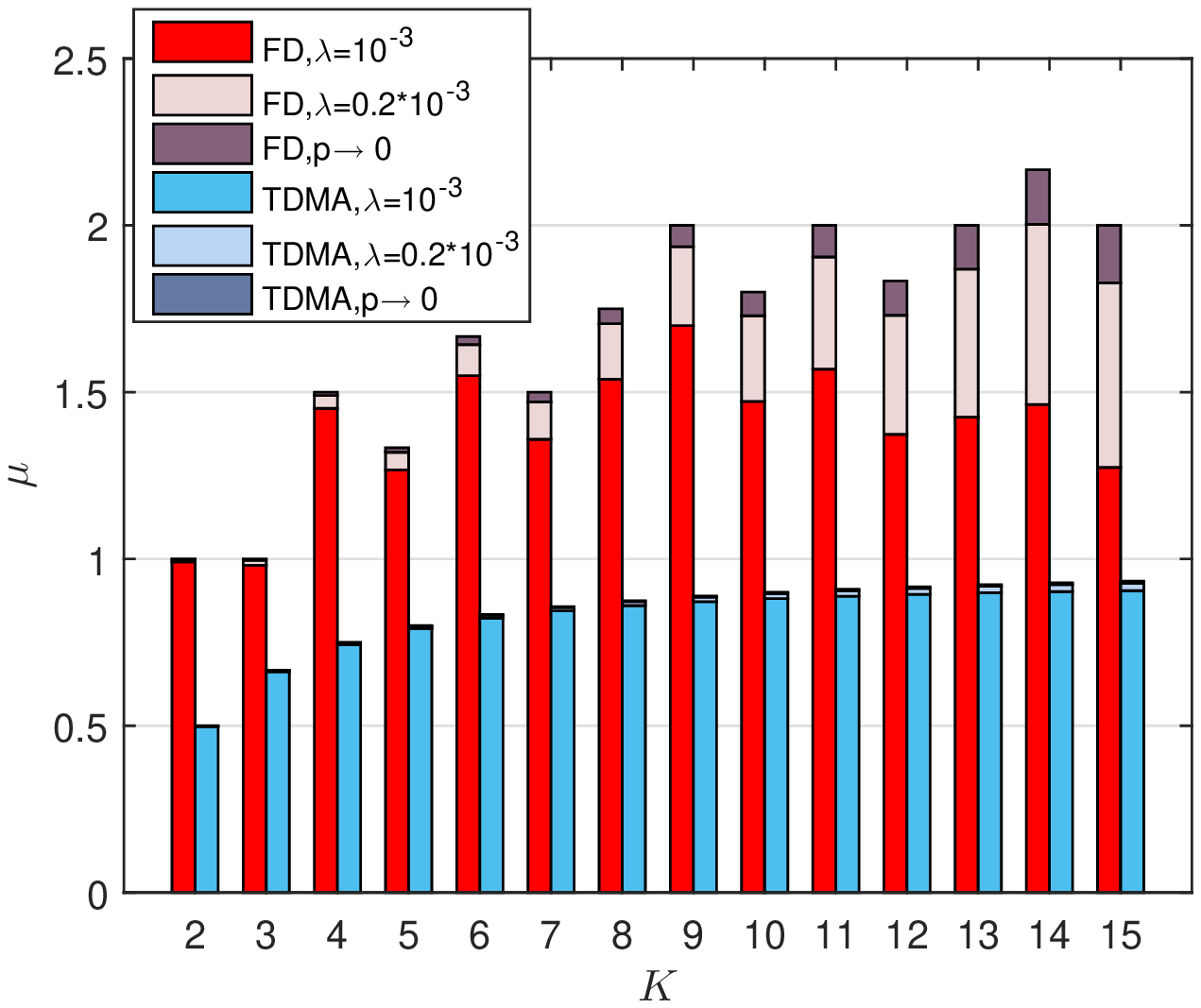}\label{muVSK_pII}}
\caption{{FD vs TDMA power talk for increasing $K$: net reception rate per unit $\mu$}.}
\label{muVSK}
\end{figure*}

Fig.~\ref{etaVSK} compares FD and TDMA power talks in terms of the achievable net transmission rates per unit, respectively, where the results in Fig.~\ref{etaVSK}\subref{etaVSK_pI} are obtained for corresponding $B_{opt}$.
Evidently, $\eta$ decreases monotonically with $K$, which could be expected.
Also, FD power talk is clearly a superior solution.
%Note that $\overline{\mu}=(K-1)\overline{\eta}$ does not increase monotonically for MW-based power talk since ${\eta}_b$ is a step function in Chang-Weldon coding for BI-MAAC; when a step in ${\eta}_b$ occurs, $\overline{\mu}$ slightly decreases.
As depicted, power talk with model change detector performs significantly better than periodic training strategy for both TDMA- and FD-based solution; the price to pay is increased implementation complexity and sensitivity to potential missed detections and false alarms, which are not included in the present study.
%However, this solution relies on the performance of the detector which in practical case is not ideal and has non-zero probability of false alarm as well as non-zero probability of missed detection.
%In general, existing model change detectors will be quite suitable for the problem of detecting voltage level change in given interval, especially when the observation noise if low enough.
%The analysis of this aspect is out of the scope here; we note however, that including the effect of system change detection imperfections into the analysis will reduce the gain of the model change detection protocol over the protocol with periodic insertion.
Further, for both protocols, the performance loss w.r.t. the stable operation, when $p\rightarrow 0$, is larger for FD-based solution due to the longer training phase.
This deviation becomes more apparent as the number of units $K$ increases, as well as for larger values of the load change intensity $\lambda$.
It can be also observed that the gain of the FD-based solution over the TDMA-based solution reduces with increasing $K$ and increasing $\lambda$.
In principle, from the equations \eqref{eff_pI_tdma}-\eqref{eff_pII_tdma}, one can determine the values of $K$ for which $\eta_{TDMA} > \eta_{FD}$, for given $\lambda$ and $B$.
%We note here that $K$ for which $\overline{\eta}_{TDMA}>\overline{\eta}_{MW}$ will be even lower if we include the overhead necessary to fight the noise related error into the analysis.
Thus, it could be concluded that FD-based power talk is more efficient for systems with smaller number of units.

We end the evaluation by reviewing the developed power talk protocols in the context of applications in which of a particular importance is the amount of information a single unit can obtain about the status of other units in the system.
Examples of such applications can be found in distributed control in MGs, average consensus and optimal dispatch.
An appropriate metric to evaluate power talk in these applications is the net reception rate, defined as the average number of bits observed by a single unit per slot, denoted with $\mu$ and calculated as $\mu=(K-1)\eta$ when the net transmission rates are equal.
Fig.~\ref{muVSK} depicts $\mu$ for both protocol implementations.
Note that for TDMA binary power talk the asymptotic upper bound for the net reception rate is 1 bit per unit per slot.
For FD variant, $\mu$ is significantly larger and increases with $K$ (although not monotonically, due to the coding rates, see Table \ref{table1}).
Thus, FD-based power talk can provide significant benefits in MG applications in which information about the status of the rest of the system is necessary.
%Moreover, small, isolated systems are nicely suited for the technique since they can be approximated with ideal system with negligible line and common bus resistances.
%Increasing the number of units that communicate simultaneously in MW-based power talk, also increase the necessary overhead to fight against the variable system state as well as the noise, making the TDMA-based power talk a viable solution in this case.
%Finally, we note that the both implementations can be combined to yield significantly improved performances which is left out for future work.

\section{Conclusions}\label{Con}

In this paper we presented power talk, a novel concept tailored for communication among units in a MicroGrid.
The core idea of power talk is to modulate information using primary control loops of the voltage source converters that regulate the bus voltage. % and modulates the information bits into subtle variations of the supplied power by each VSC unit.
%We introduce the notions of signaling space where power talk symbol constellations can be designed and detection space where modulation and demodulation of the symbols is performed.
We have shown that it is possible to design signaling constellations that conform to the operating constraints and limits to the power deviations with respect to the nominal operation.
We have also shown that using MAP detector at the receiving side practically achieves errorless communication when the load (i.e., power demand) is stable, under mild constraints on the number of units in the system and allowable power deviations.

The main challenge of power talk is the arbitrary variations of load, leading to the uncontrollable and unforeseeable changes of the bus voltage.
We investigated techniques to counter-effect load changes, showing that it is possible to optimize the power talk operation given the statistics of the load changes.

The achievable rates of power talk depend on the bandwidth of the primary control loops.
In practice, it could be expected that power talk can achieve rates of the order of $100 \, \text{Baud} - 1 \, \text{kBaud}$.
Nevertheless, considering that the inter-MG communications are machine-type in nature, these modest rates may prove to be satisfactory.
Moreover, when assessing the potential of power talk, one should also take into account the inherent advantages of power talk, which are use of existing MG power equipment, software implementation, and reliability and availability equal to the reliability and availability of the MG itself.

\appendices

\section{Analysis of power talk with periodic insertion of training sequences}\label{appA}

\subsection{TDMA power talk}\label{pITDMA}

%With TDMA, no coding is done to communicate in multiple access settings.
When scheduled to transmit, VSC $k$ transmits exactly $1$ bit of information in a single slot with probability $1-p$, {where $p$ is the probability that the load changes during the slot, see Section~\ref{Protocols}.}
Under fair scheduling and in the absence of noise, the transmission rates are equal for all units.
Denote with $t,1\leq t\leq KB$ the slot when the first change of the system state occurs.
Then, $\eta_{\text{FD}}$ can be written as:
\begin{equation}\label{pItdma1}
\eta_{\text{TDMA}}=\frac{(1-p)^{L_{\text{TDMA}}}}{L_{\text{TDMA}}+KB}\bigg[\frac{p}{K}\sum_{t=1}^{KB}(t-1)(1-p)^{t-1}+B(1-p)^{KB}\bigg].
\end{equation}
The last term corresponds to the case when no load change occurs during the data phase.
Using arithmetic-geometric progression to solve \eqref{pItdma1} produces \eqref{eff_pI_tdma}.

\subsection{FD power talk}\label{pIMW}
With FD, we use Chang-Weldon uniquely decodable coding and decoding \cite{ref:28}.
The code for VSC $k$ contains only two codewords of length $\frac{1}{\eta_{\text{FD}}^\texttt{S}}$ to represent each bit, with $\eta_{\text{FD}}^\texttt{S}$ given in Table II.
Thus, when using FD solution, each unit has to send a block of bits of length $\frac{1}{\eta_{\text{FD}}^\texttt{S}}$ correctly to be deliver 1 bit of information that can be uniquely decoded by other units.
Again, in absence of noise the transmission rates are equal for all units.
Denote with $\tau,1\leq\tau\leq B$ the block of bits of length $\frac{1}{\eta_{\text{FD}}^\texttt{S}}$ when the first load change occurs.
Then, $\eta_{\text{FD}}$ can be written as:
\begin{equation}\label{pIMW}
\eta_{\text{FD}}=\frac{(1-p)^{L_{\text{FD}}}}{L_{\text{FD}}+\frac{B}{\eta_{\text{FD}}^\texttt{S}}}\bigg[\sum_{\tau=1}^{B}(\tau-1)(1-p)^{\frac{\tau-1}{\eta_{\text{FD}}^\texttt{S}}}(1-(1-p)^{\frac{1}{\eta_{\text{FD}}^\texttt{S}}})+B(1-p)^{\frac{B}{\eta_{\text{FD}}^\texttt{S}}}\bigg].
\end{equation}
Using arithmetic-geometric progression to solve \eqref{pIMW}, produces \eqref{eff_pI_mw}.

\section{Analysis of power talk with load change tracker}\label{appB}
\begin{figure}[h]
\centering
\includegraphics[scale=0.2]{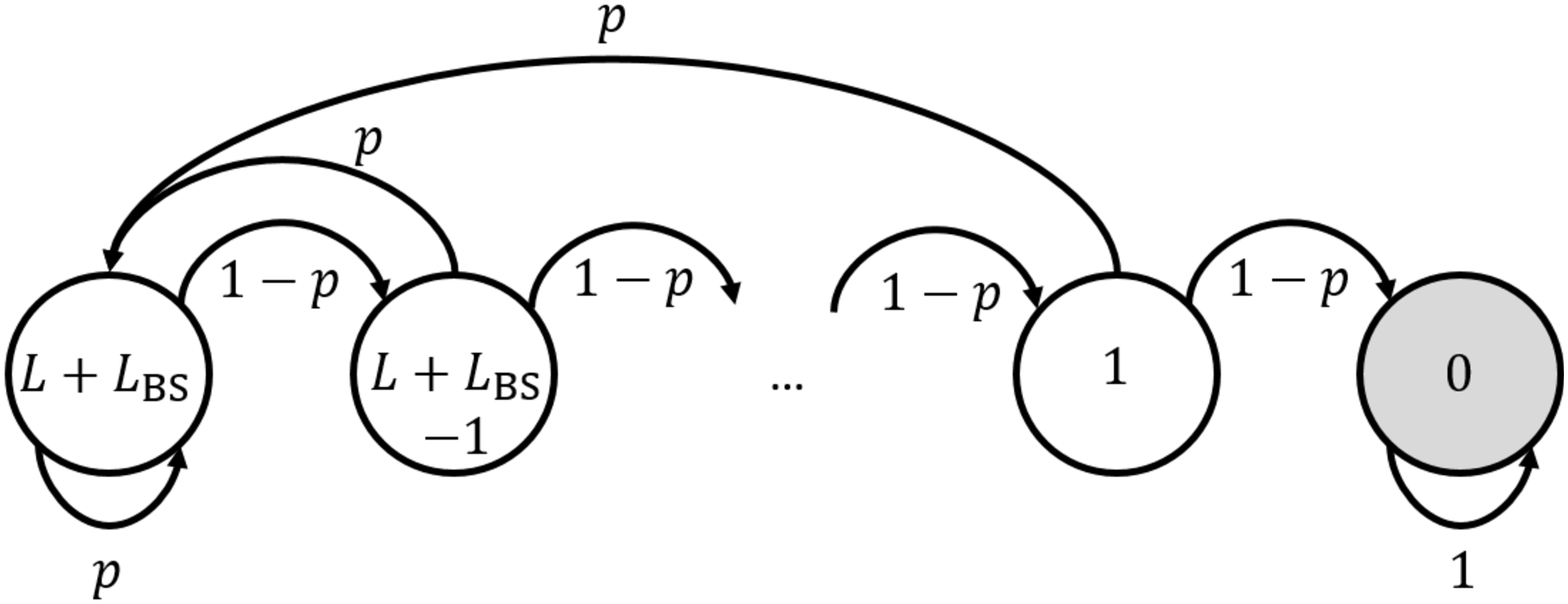}
\caption{Absorbing Markov chain as a model for $L^{'}$}
\label{chain}
\end{figure}

The derivation is identical for both TDMA- and FD-based power talk.
The average number of slots, necessary to deliver $B$ bits of information depends on the number of state changes that occur during the transmission of $B$ bits.
In absence of noise, the average number of slots, necessary to deliver $B$ bits is the same for all units and $\eta$ can be written as:
\begin{equation}\label{ave_time}
\eta=\frac{B}{\frac{B}{\eta^\texttt{S}}+p\frac{B}{\eta^\texttt{S}}(\mathbb{E}\left\{L^{'}\right\}+1)}=\frac{1}{1+p(\mathbb{E}\left\{L^{'}\right\}+1)}\eta^\texttt{S},
\end{equation}
and it does not depend on $B$ and $\mathbb{E}\left\{L^{'}\right\}$ is the average duration of the training phase, since the state can also change during the training.
We also include the slot in which the state changed, since the bit in that slot will be retransmitted.
To analyze $\mathbb{E}\left\{L^{'}\right\}$ we note that if a load change occurs in the training sequence, the training sequence is re initiated.
Then, we use the absorbing Markov chain shown on Fig.~\ref{chain} to model $L^{'}$.
The chain always starts from the state $L+L_{BS}$.
The absorbing state is the state when the training sequence has been completely sent without interruptions.
$\mathbb{E}\left\{L^{'}\right\}$ is actually the average number of transitions in the chain until absorption.
%The transition matrix of the chain can be written in the canonical form as follows:
%\begin{equation}\label{trans}
%\mathbf{P}=
%\left[
%\begin{array}{ccccc|c}
%p & 1-p & 0 &...& 0 & 0 \\ 
%p & 0 & 1-p &...& 0 & 0 \\
%\vdots & \vdots & \vdots & & \vdots & \vdots \\
%p & 0 & 0 &...& 0 & 1-p \\\hline
%0 & 0 & 0 &...& 0 & 1
%\end{array}\right]
%=\left[
%\begin{array}{c|c}
%\mathbf{Q} & \mathbf{R} \\\hline
%\mathbf{0} & \mathbf{I}
%\end{array}\right].
%\end{equation}
%Then, the average number of transitions to absorption if the chain starts at state $L+L_{SS}$ is the first element of the vector $(\mathbf{I}-\mathbf{Q})^{-1}\mathbf{1}$ where $\mathbf{1}$ is the all ones vector. 
For the chain shown of Fig.~\ref{chain}, it can be shown that the average number of transitions until absorption is:
\begin{equation}\label{absorb}
\mathbb{E}\left\{L^{'}\right\}=\sum_{l=1}^{L+L_{BS}}(1-p)^{-l}.
\end{equation}
Replacing \eqref{absorb} in \eqref{ave_time} and solving the geometric sum, produces \eqref{eff_pII_tdma}.

%\section*{Acknowledgment}

%\nocite{*}
%\bibliographystyle{IEEEtran}
%\bibliography{IEEEabrv,PT}

\end{document}